%% file: nrg-review.tex
\documentclass[11pt]{article}
\pdfoutput=1

\usepackage[usenames,dvipsnames]{xcolor}
\usepackage[a4paper]{geometry}
\usepackage[english]{babel}
\usepackage{cite,enumerate,enumitem,booktabs,float,graphicx}

\usepackage[T1]{fontenc}
\usepackage[utf8]{inputenc}
\usepackage{lmodern}

\makeatletter
\g@addto@macro\bfseries{\boldmath}
\makeatother

\usepackage{bm,amsmath,amssymb,tensor,mathtools}
\numberwithin{equation}{section}
\usepackage{nth}

\usepackage{tikz}
\usetikzlibrary{cd}

\usepackage{caption}
\usepackage{subcaption}

\usepackage[pdftex]{hyperref}
\definecolor{dark-blue}{rgb}{0.15,0.15,0.4}
\hypersetup{
  colorlinks,
  linkcolor={dark-blue},
  citecolor={blue}
}

\usepackage{enumitem}
\usepackage{amsthm}

\usepackage{authblk}

\input{defs.tex}

\newcommand{\spa}{\ \ ,\ \ \ \ }

\usepackage{amssymb}
\usepackage{booktabs}
\usepackage{cancel}
\usepackage{setspace}
\usepackage{authblk}
\usepackage{delimset}
\usepackage{accents}
\usepackage{tikz}
\usetikzlibrary{matrix}
\numberwithin{equation}{section}

\newcommand{\order}[1]{\mathcal{O}\brk{#1}}
\newcommand{\os}[2]{\accentset{\scriptscriptstyle{\brk{\hspace{-0.05em}#2\hspace{-0.05em}}}}{#1}{}}

\newcommand{\oss}[3]{{\os{#1}{#2}}_{\mathrm{#3}}}

\title{\textbf{%
Review on Non-Relativistic Gravity}
}
\date{}

\author[1]{Jelle Hartong}
\author[2,3]{Niels A. Obers}
\author[3]{Gerben Oling}
\affil[1]{%
  School of Mathematics and Maxwell Institute for Mathematical Sciences,
  \protect\\
  University of Edinburgh,
  Peter Guthrie Tait Road,
  Edinburgh EH9 3FD, UK
}
\affil[2]{%
  The Niels Bohr Institute,
  University of Copenhagen,
  \protect\\
  Blegdamsvej 17,
  DK-2100 Copenhagen Ø,
  Denmark
}
\affil[3]{%
  Nordita,
  KTH Royal Institute of Technology and Stockholm University,
  \protect\\
  Hannes Alfvéns väg 12, SE-106 91 Stockholm, Sweden
}

\begin{document}

\maketitle
\thispagestyle{empty}

\begin{abstract}
\noindent
We review the history of Newton--Cartan gravity with an emphasis on recent developments, including the covariant, off-shell large speed of light expansion of general relativity.
Depending on the matter content, this expansion either leads to Newton--Cartan geometry with absolute time or to Newton--Cartan geometry with non-relativistic gravitational time dilation effects.
The latter shows that non-relativistic gravity includes a strong field regime and goes beyond Newtonian gravity.
We start by reviewing early developments in Newton--Cartan geometry, including the covariant description of Newtonian gravity, mainly through the works of Trautman, Dautcourt, K\"unzle and Ehlers.
We then turn to more modern developments, such as the gauging of the Bargmann algebra, and we describe why the latter cannot be used to find an off-shell covariant description of Newtonian gravity.
We review recent work on the $1/c$ expansion of general relativity and show that this leads to an alternative `type II' notion of Newton--Cartan geometry.
Finally, we discuss matter couplings, solutions and odd powers in $1/c$, and we conclude with a brief summary of related topics.
\end{abstract}

\newpage
\tableofcontents

\section{Introduction}
While at the fundamental level Nature is Lorentz invariant, there are many
instances where it appears effectively non-relativistic. 
On the one hand, this can happen in  condensed matter or biological systems, but it can obviously also be true for gravitational phenomena. Indeed, it is well known that Newtonian gravity can
be obtained as a limit of general relativity.
Additionally, general relativity can often be approximated using non-relativistic descriptions of gravity, such as in the post-Newtonian approximation.
This review of non-relativistic gravity is based on a modern description of non-relativistic approximations in terms of Newton--Cartan geometries, their dynamics and their interaction with matter.

The idea of geometrizing Newton's law of gravitation
dates back all the way to the introduction of Newton--Cartan geometry by Cartan in 1923~\cite{Cartan1,Cartan2}. 
The deep insight at the heart of that work is that ``gravity is geometry''
is true independently of whether local observers in inertial frames see the laws of special relativity or of Galilean relativity.
Special relativity in inertial frames is a separate and essential ingredient in the formulation of Einstein's theory of general relativity.
On the other hand, inertial frames with local Galilean relativity lead to Newton--Cartan gravity.

Recent years have seen a revival of research on non-relativistic gravity.
In particular, two important novel insights have been obtained.
First of all, we have learned that non-relativistic gravity is much richer than Newtonian gravity, and goes beyond it by allowing for a strong field regime, including gravitational time dilation.
One key ingredient for this was allowing for Newton--Cartan metric-compatible connections with nonzero torsion.
Geometrically, nonzero torsion leads to spacetime manifolds without absolute time, which goes beyond the historical perspective on Newton--Cartan (NC) geometry where time is always absolute.
Another crucial development came from considering the off-shell large speed of light expansions, as opposed to on-shell expansions or large speed of light limits.
These off-shell expansions naturally led to a notion of NC geometry whose local symmetries are different than what was previously considered, which turns out to be necessary to provide an action principle for non-relativistic gravity and Newtonian gravity in particular.

The recent revival of non-relativistic gravity has been spurred on in large part by 
a deeper understanding of non-relativistic geometry and its multiple connections to field theory, holography and string theory.%
\footnote{%
  In addition to this Introduction, many more relevant and important works will be mentioned in the following sections and the final discussion section.
  We also refer the reader to the companion reviews on other topics in applications of non-relativistic (and more generally non-Lorentzian) geometry~\cite{Grosvenor:2021hkn,Oling:2022fft,Bergshoeff:2022eog}.
}
We highlight here in particular the 2010 paper~\cite{Andringa:2010it} by Andringa, Bergshoeff, Panda and De Roo, which showed how to obtain torsionless NC geometry by gauging the Bargmann algebra (which is the Galilei algebra with a central extension).
This provided a modern understanding of the geometric fields of NC geometry. 
Another important advance was the discovery of a torsionful generalization
of NC geometry, which was first observed as the boundary geometry in the context of non-AdS (Lifshitz) holography \cite{Christensen:2013lma,Christensen:2013rfa,Hartong:2014pma}.
In a remarkable parallel development, Son showed in \cite{Son:2013rqa} that the non-relativistic effective field theory describing certain aspects of the fractional quantum Hall effect naturally couples to NC geometry, highlighting its relevance for non-relativistic field theories.
Finally, a systematic analysis of the large speed of light expansion of GR, taking into account the possibility of torsion in NC geometry, was done in Refs~\cite{VandenBleeken:2017rij,Hansen:2018ofj,Hansen:2020pqs,Ergen:2020yop}, leading to our present understanding of non-relativistic gravity.

We start this review in Section~\ref{sec:history} by presenting an overview of the most pertinent advances in the historical development of NC geometry up to the recent revival of NC geometry ten years ago.
We will not be fully exhaustive, but we have attempted to collect the main references and milestones into a readable introduction to the subject of (torsion-free) NC geometry.
Further details are given in Section~\ref{sec:nc-basics},
where we present the notion of Newton--Cartan geometry
that was developed in the work of Trautman, Dautcourt, K\"unzle, Ehlers and others. 
In contrast to some of these original works, our discussion here is fully covariant.

Then in Section~\ref{sec:recenthistory} we turn to the modern era, which we define as starting from the aforementioned paper~\cite{Andringa:2010it}, where it was shown that NC gravity with absolute time can be viewed as the
dynamics of a geometry that can be obtained by gauging the Bargmann algebra.
Details of this procedure, including its extension to nonzero torsion, are discussed in Section~\ref{sec:type-I-tnc}.
This results in what we refer to as `type I' torsional Newton--Cartan (TNC) geometry.
We obtain the action for a non-relativistic point particle coupled to this geometry, as well as an action for its gravitational dynamics.
On a technical level, these actions are obtained from two distinct routes,
namely a large speed of light limit and null reduction,
whose results turn out to be equivalent.
At the end of this Section, we show that type I TNC is
not the correct geometrical framework to get an off-shell action for Newtonian gravity. 

We thus turn in Section~\ref{sec:NRexpGR} to the derivation of an action which encapsulates the Poisson equation of Newtonian gravity as an equation of motion.
In fact, this action describes a more general notion of non-relativistic gravity which includes a strong field regime.
This action is obtained by carefully considering the large speed of light expansion of GR, which naturally leads to a new notion of Newton–Cartan geometry, which we refer to as `type II' TNC geometry. 

Section~\ref{sec:aspectsNRG} then briefly discusses various other aspects of non-relativistic gravity from the perspective of the large speed of light expansion, including the coupling to matter and the particular case of the strong field expansion of the Schwarzschild solution of GR, as well as some other examples of solutions.
We also remark on the inclusion of odd powers of $1/c$ in the expansion, as the analysis in Section~\ref{sec:NRexpGR} is restricted to even powers, which form a closed subsector.

Finally, Section~\ref{sec:discussion} contains a discussion of various related applications and appearances of non-relativistic gravity in the fast-growing recent literature on these and related topics, including many additional references.

\section{History}\label{sec:history}
We start by giving an overview of some of the important historical developments.
We will not attempt to be exhaustive, and it is not our aim to present each individual paper's contents as accurately as possible.
Rather, we want to weave an accessible narrative through the early developments of the subject. 
This section and the next can also be read as an introduction to the subject of Newton--Cartan geometry.
We have made an effort to provide clickable links to the relevant papers, which can be hard to find.

Historically, the subject was introduced by Cartan in~\cite{Cartan1,Cartan2}; see~\cite{Ashtekar} for an English translation.
However, these papers are not easily digestible as they are not written in a modern geometrical language,
so it would
be difficult to incorporate them accurately in our narrative.
For this reason, and with much regret, we will not have anything to say about these works.

Instead, the earliest historical reference for our current discussion and one of the pioneering works on the subject of $1/c$ expansions of general relativity is the paper by Friedrichs \cite{Friedrichs}.
This paper introduces the Newton--Cartan metric $\tau_\mu\tau_\nu$ and co-metric $h^{\mu\nu}$, which is symmetric and has signature $(0,1,\ldots,1)$, and writes Newton's second law in a covariant form in terms of a Newton--Cartan metric-compatible connection, while realizing that the latter is not unique.
The paper then finds these objects from general relativity (GR) via a $1/c$ expansion.
Similar comments about Newton--Cartan metrics and their properties can be found in older work by Weyl~\cite{Weyl}.

For an overview of early aspects of Newton--Cartan geometry we refer the reader to the review by Havas~\cite{Havas}.
The subject has also been covered in the books by Malament~\cite{Malament} and Misner, Thorne and Wheeler~\cite{Misner:1973prb}.

\subsection{Trautman, Dautcourt, K\"unzle, Ehlers}
Despite these important pioneering works, we consider the more modern notion of Newton--Cartan geometry to begin in earnest with the work%
\footnote{%
  The works of Trautman can be found here: \href{http://trautman.fuw.edu.pl/publications/scientific-articles.html}{http://trautman.fuw.edu.pl/publications/scientific-articles.html}.
 The first reference \cite{Trautman1} is in French.
 The second reference \cite{Trautman2} is a chapter from a book (see in particular Section 5.3) and it basically reviews the contents of \cite{Trautman1} with slightly more detail.
 The later paper \cite{Trautman3} is essentially an English version of \cite{Trautman1}.
}
of Trautman~\cite{Trautman1,Trautman2,Trautman3}, which provides an axiomatic definition of Newtonian gravity.

In all three papers Trautman gives an axiomatic definition of Newton--Cartan gravity.
The precise set of postulates changes slightly from paper to paper.
We follow \cite{Trautman3}, where Trautman introduces the following postulates: 
\begin{enumerate}
  \item
    \label{it:traut1-smooth-spacetime}
    Spacetime is a four-dimensional differentiable manifold endowed with a symmetric affine connection $\Gamma^\rho_{\mu\nu}$.
  \item
    \label{it:traut2-deg-metrics}
    There is a nowhere-vanishing one-form $\tau_\mu$ and a nowhere vanishing co-metric $h^{\mu\nu}$ of signature $(0,1,1,1)$ such that $h^{\mu\nu}\tau_\mu=0$.
  \item
    \label{it:traut3-metric-compatibility-symmetric}
    The symmetric affine connection $\Gamma^\rho_{\mu\nu}$ is metric-compatible in the sense that 
    \begin{equation}
      \nabla_\mu\tau_\nu=0\,,
      \qquad
      \nabla_\mu h^{\nu\rho}=0\,,
    \end{equation}
    where $\nabla_\mu$ is the associated covariant derivative.
  \item
    \label{it:traut4-curvature-conditions}
    The Riemann curvature tensor $R_{\mu\nu\rho}{}^\sigma$ associated with the affine connection obeys the following two conditions:
    \begin{equation}\label{Trautman1}
      R_{\kappa\lambda[\mu}{}^\rho\tau_{\nu]}=0\,,
      \qquad R_{\mu}{}^\nu{}_\rho{}^\sigma=R_{\rho}{}^\sigma{}_\mu{}^\nu\,,
    \end{equation}
    where the second index has been raised with $h^{\mu\nu}$,
    so $R_{\mu}{}^\nu{}_\rho{}^\sigma=h^{\nu\kappa}R_{\mu\kappa}{}_\rho{}^\sigma$.
    See Appendix~\ref{sec:conventions}
    for our conventions for the Riemann tensor
  \item
    \label{it:traut5-covariant-poisson}
    Particles in free fall follow geodesics of the affine connection and gravity is described by an equation of the form (in $3+1$ dimensions)
    \begin{equation}\label{Trautman3}
      R_{\mu\nu}=4\pi G\rho\tau_\mu\tau_\nu\,,
    \end{equation}
    where $\rho$ is the mass density, $G$ is Newton's constant and $R_{\mu\nu}=R_{\mu\rho\nu}{}^\rho$ is the Ricci tensor.
\end{enumerate}
Since the connection is symmetric, taking the antisymmetric part of $\nabla_\mu\tau_\nu=0$ implies that $2\pd_{[\mu}\tau_{\nu]}=0$, or equivalently $d\tau=0$ in form notation.
The first of the two conditions in \eqref{Trautman1} has later been dropped in favour of a boundary condition, as we will discuss below.
The second condition in~\eqref{Trautman1} is often simply referred to as the Trautman condition.
In various places in the literature, it is phrased as
\begin{equation}\label{Trautman2}
  h^{\lambda[\mu}R_{\lambda(\rho\sigma)}{}^{\nu]}=0\,.
\end{equation}
The latter is equivalent to 
\begin{equation}
  R_{(\mu}{}^\nu{}_{\rho)}{}^\sigma=R_{(\mu}{}^\sigma{}_{\rho)}{}^\nu\,,
\end{equation}
which appears to be a weaker condition than the second condition of~\eqref{Trautman1}.
However, using $\nabla_{\rho}h^{\mu\nu}=0$, which implies
$[\nabla_{\rho},\nabla_{\sigma}]h^{\mu\nu}=0$
or equivalently
$h^{\kappa\nu}R_{\rho\sigma\kappa}{}^\mu+h^{\kappa\mu}R_{\rho\sigma\kappa}{}^\nu=0$,
equation~\eqref{Trautman2} can be shown to be equivalent to the second condition of~\eqref{Trautman1}.

Around the time of Trautman's work, the mathematical framework underlying Newton--Cartan geometry was put on a firm foundation in the work of Dombrowski and Horneffer~\cite{Dombrowski} (see also the later work by K\"unzle \cite{Kuenzle1}).
We will not focus on the more mathematical development of Newton--Cartan geometry in this review, so we will not comment further on these works. 

Trautman's work can be viewed as an intrinsic definition of Newtonian gravity without any reference to $1/c$ expansions of GR.
It was later improved by K\"unzle \cite{Kuenzle1}, who among other things worked out the class of torsion-free metric compatible connections and realized that the first Trautman condition 
(the first equation in~\eqref{Trautman1})
is redundant if we assume asymptotic flatness.
It was shown by Dautcourt~\cite{Dautcourt1,Dautcourt2} that an appropriate covariant expansion of GR in powers of $1/c$ reproduces Trautman's formulation of Newton--Cartan gravity, with the caveat that the first Trautman condition
is dropped.
We refer the reader to Sections \ref{subsec:1/cGR} and \ref{sec:NRexpGR}
for more details about this.
The first Trautman condition is equivalent to the condition $h^{\rho\kappa}R_{\mu\nu\rho}{}^\sigma=0$, which Dautcourt in \cite{Dautcourt2} attributes to Dixon%
\footnote{%
  In \cite{Dixon} it is shown that Newtonian gravity, as defined under point 5 of Trautman's axiomatic definition (plus a cosmological constant) is unique, in that it follows from points 1 to 3 as well as the second equation of \eqref{Trautman1} and some general assumptions about the structure of the equation and the matter content.
}
\cite{Dixon}.
Dautcourt writes that this condition does not follow from the $1/c$ expansion of GR.
On the other hand, the second equation of~\eqref{Trautman1} does follow from the $1/c$ expansion of the Riemann tensor of GR.
More details will be given below.

In \cite{Dautcourt2}, Dautcourt only discusses the expansion in terms of even powers of $1/c$.
From the outset, the Newton--Cartan connection is taken to be symmetric, and metric compatible.
As we saw above, this implies that $d\tau=0$ so that%
\footnote{%
  We do not allow for non-contractible closed timelike loops, so that $\tau$ is exact when it is closed.
}
$\tau=dT$ for a time coordinate $T$.
Therefore, in this notion of NC geometry, time is absolute.
This is an important restriction from the $1/c$ expansion perspective.
Dautcourt is aware that the latter can be formulated without this restriction
but does not consider the general case.
 
In a later paper~\cite{Dautcourt:1996pm}, Dautcourt includes odd powers in $1/c$.
This paper more explicitly discusses the option of allowing for a nontrivial lapse function $N$ in NC geometry, so that $\tau=NdT$, which is now known as twistless torsional NC geometry (TTNC).
It is equivalent to the Frobenius condition $\tau\wedge d\tau=0$ implying that equal $T$ surfaces define spatial hypersurfaces.%
\footnote{%
  As we will see in Section~\ref{sec:NRexpGR}, the $1/c$ expansion of Einstein's equations rules out the possibility that $\tau\wedge d\tau\neq 0$.
}
In particular, the paper concludes that the $1/c$ expansion of GR leads to a theory that is more general than Newtonian gravity (on page 7), but then suggests that this more general case with $N$ a priori arbitrary
is not so interesting because insisting on global regularity of the Newton--Cartan lapse function $N$ would reduce $N$ to a constant and would thus force time to be absolute.
The argument here is that the NC lapse function is harmonic and, by Liouville's theorem, it must therefore be constant in order to be regular everywhere without appropriate sources.
The current viewpoint is that the condition of global regularity of the NC lapse function is too restrictive,
as this rules out interesting non-relativistic approximations of GR (see further below), and that one can find sources leading to singularities in $N$.

In \cite{Ehlers1,Ehlers2} Ehlers develops `frame theory', which is a geometrical formulation that treats Lorentzian and Galilean geometry on equal footing.
Frame theory has a parameter $\lambda=c^{-2}$ that can be either zero or positive, leading to Galilean or Lorentzian geometry.
Furthermore, it has a second parameter, Newton's constant $G$, that can also be zero or positive, leading to four different theories depending on whether $\lambda$ and $G$ are zero or positive.
Setting $G=0$ means that one is considering test particles in a fixed background that obeys some curvature constraint.
The motivation behind frame theory is to put the $1/c$ expansion on a firmer mathematical foundation, improving on the work of Dautcourt.
For a review on frame theory we refer the reader to \cite{Oliynyk:2009pz}.

Ehlers reviews Trautman's work and notices that the condition in the first equation of \eqref{Trautman1} can be written in other forms.
In particular, it can be shown that the following equations are all equivalent formulations of this condition:
\begin{eqnarray}
   0 & = & R_{\mu\nu\sigma}{}^\rho R_{\alpha\beta\rho}{}^\sigma h^{\nu\beta}\,,\\
   0 & = &  R_{\mu\nu[\sigma}{}^\rho\tau_{\gamma]}\,,\\
   0 & = & R_{\mu\nu\sigma}{}^{[\rho}h^{\gamma]\sigma}\,.
\end{eqnarray}
These three conditions are now often referred to as the Ehlers conditions.
The main point of these conditions is to restrict the geometry such that the Newton--Cartan gravity equation \eqref{Trautman3} only contains the Newtonian potential (in appropriate coordinates).
However, as we know from Dautcourt's work (and as was also known to Ehlers) these conditions are unnecessary, as this restriction can also be achieved by imposing asymptotic flatness.
In \cite{Ehlers3}, Ehlers gives a short review of frame theory and gives examples of Newtonian limits of well-known GR solutions.

\subsection{Post-Newtonian corrections}
Part of the motivation behind frame theory is the question whether or not solutions to Newtonian gravity are extendable
(for example in the sense of post-Newtonian corrections)
to full relativistic solutions of GR.
In a paper by Rendall \cite{Rendall} it is shown that post-Newtonian corrections are not compatible with asymptotic flatness.
The physical reason behind this is that these corrections do not correctly describe the far zone of some non-relativistic matter distribution where gravitational waves dominate.
In other words, the post-Newtonian regime corresponding to for example a perfect fluid matter source has a finite radius of validity.
This is an effect that is noticeable when one goes beyond the first post-Newtonian correction.
In \cite{Lottermoser1,Lottermoser2}, Lottermoser showed the constraint equations of GR (in harmonic gauge) admit a well-defined $1/c$ expansion (in the sense of a convergent series).

In current literature on post-Newtonian corrections, which will not be reviewed here (see for example \cite{WillPoisson,Blanchet:2004ek} and references), the dominant approach does not consist of expanding the Einstein equations in powers of $1/c$ and then solving them order by order, which is sometimes called the classic approach.
Instead, one formally solves the Einstein equations in harmonic gauge (using what are known as the relaxed Einstein equations%
\footnote{%
  The relaxed Einstein equations are equivalent to Einstein's equations but written in terms of a different variable
  $k^{\mu\nu}=\sqrt{-g}g^{\mu\nu}-\eta^{\mu\nu}$ where $\eta_{\mu\nu}$ is the Minkowski metric.
  Furthermore, this formulation only works in harmonic gauge for which $\partial_\mu k^{\mu\nu}=0$.
})
and imposes a boundary condition that leads to an integral equation using a retarded Green's function.
In the Blanchet--Damour approach%
\footnote{%
See~\cite{Blanchet:2004ek} and references therein.
An alternative approach to solving this integral equation is the Will--Wiseman approach, see~\cite{WillPoisson} and references therein.
},
this integral equation is solved outside the source (i.e. in vacuum) as an expansion in $G$.
In a region containing the non-relativistic source, the integral equation is solved as an expansion in $1/c$.
The $G$ and $1/c$ expansions are then matched multipole by multipole in their overlap region using matched asymptotic expansion.

The recent work on the covariant $1/c$ expansion of GR can be viewed as an attempt to revive the classic approach used in the early days of work on post-Newtonian expansions.
It also goes beyond that because it can cover regimes of strong gravity%
\footnote{%
  By strong gravity we mean a regime where the clock 1-form $\tau$ obeys $\tau\wedge d\tau=0$ but $d\tau\neq 0$ so that there is a nontrivial NC lapse function (describing a NC geometry with gravitational time dilation).
}
and is more flexible as to the gauge choice one uses.
In general, one will have to match the $1/c$ expansion onto a $G$ expansion, and thus find some hybrid of the classical and more modern Blanchet--Damour or Will-Wiseman approaches, which is the aim of the upcoming work \cite{HartongMusaeus}.
(See also \cite{Tichy:2011te} for a covariant approach to the post-Newtonian expansion up to 1PN order).

\subsection{Null reduction \label{sec:nullred}}
Finally, even though it is not the focus of this review paper, we would be remiss not to mention the work by Duval, Burdet, Künzle and Perrin~\cite{Kuenzle3}.
So far, we have discussed Newton--Cartan geometry either intrinsically or as originating from the $1/c$ expansion of GR.
Duval and collaborators offered a third perspective by showing that Newton--Cartan geometry can also be obtained from null reduction of a Lorentzian geometry with a null isometry.
The paper \cite{Kuenzle3} (see also \cite{Duval:1990hj} for further developments and~\cite{Duval:2009vt} for a review) considers null reduction of Lorentzian geometry and null uplifts of Newton--Cartan geometry to Lorentzian geometry with a null Killing vector. 
This is related to the Eisenhart lift \cite{Eisenhart} of Hamiltonian dynamics as shown in~\cite{Minguzzi:2006gq}.%
\footnote{%
  See for example the review paper~\cite{Cariglia:2015bla} for details about the Lorentzian Eisenhart lift.
}
It is also related to the Bargmann algebra \cite{Bargmann:1954gh}, since this algebra is the centralizer of the null isometry in the higher-dimensional Poincaré algebra~\cite{Gomis:1978mv}, as we will review around Equation~\eqref{eq:null-centralizer-generators}.

The paper \cite{Kuenzle3} discusses uplifts of Newton--Cartan geometries for which $d\tau=0$ to pp-waves in one dimension higher. This can be generalised to cases where $\tau\wedge d\tau=0$ \cite{Julia:1994bs} and even to cases where $\tau\wedge d\tau\neq 0$ \cite{Hartong:2014oma}.

However, as we will review in Section~\ref{ssec:type-I-tnc-mass-coupling}, the equation~\eqref{Trautman3} for Newton--Cartan gravity coupled to matter cannot be obtained from null reduction.
Despite this shortcoming,
it is still useful to adapt this higher-dimensional perspective
for many aspects such as particle motion in a fixed background as well as more geometrical questions.

\section{Basics of torsion-free Newton--Cartan geometry}
\label{sec:nc-basics}

This section provides more details about the covariant formulation of Newton--Cartan geometry that started with the work of Trautman.
This section does not follow a strict historical path and uses slightly more modern tools.
For example, in contrast to much of the early literature we will use a fully covariant approach and largely refrain from choosing special coordinates. 

\subsection{Newton--Cartan metric data}
The main objects of interest are the nowhere-vanishing `clock' one-form $\tau_\mu$ and the nowhere-vanishing `spatial' co-metric $h^{\mu\nu}$ which has signature $(0,1,\ldots, 1)$.
These objects obey the condition that $\tau_\mu h^{\mu\nu}=0$.
We take the dimension of the underlying spacetime manifold to be $(d+1)$, so that the spatial co-metric has rank $d$.
We can also define the inverse objects $v^\mu$ and $h_{\mu\nu}$ by demanding that the following relations hold:
\begin{equation}
  \label{eq:first-nc-orthonormality}
  h_{\mu\rho}h^{\rho\nu}-\tau_\mu v^\nu=\delta^\nu_\mu\,,
  \qquad
  v^\mu h_{\mu\nu}=0\,,
  \qquad
  \tau_\mu v^\mu=-1\,,
  \qquad
  \tau_\mu h^{\mu\nu} = 0\,.
\end{equation} 
Another way of phrasing these relations among the various objects is by saying that $\tau_\mu\tau_\nu+h_{\mu\nu}$ is invertible with inverse $v^{\nu}v^\rho+h^{\nu\rho}$. The positive determinant of $\tau_\mu\tau_\nu+h_{\mu\nu}$ is denoted by $e^2$, and we can use $e$ as an integration measure.
The objects $v^\mu$ and $h_{\mu\nu}$ are defined up to a local Galilean boost, which acts as 
\begin{equation}
\delta v^\mu=h^{\mu\nu}\lambda_\nu\,,\qquad \delta h_{\mu\nu}=\tau_\mu\lambda_\nu+\tau_\nu\lambda_\mu\,.
\end{equation}
Here, we require $v^\mu\lambda_\mu=0$,
and $\lambda_\mu$ transforms under a second Galilean boost transformation as $\delta\lambda_\mu=\tau_\mu h^{\nu\rho}\lambda_\nu\lambda_\rho$ in order that $v^\mu\lambda_\mu=0$ and $v^\mu h_{\mu\nu}=0$ are boost invariant under a second order boost transformation.
The exponentiation of the infinitesimal Galilean boost transformation terminates after the second order in $\lambda_\mu$. The reason why we call these transformations Galilean boosts\footnote{They are also sometimes referred to as Milne boosts.} will become clear in Section \ref{sec:type-I-tnc}. We note that the integration measure $e$ is Galilean boost invariant.
Finally, we can introduce frame fields $e_\mu{}^a$ and $e^\mu{}_a$ for the degenerate spatial metric and co-metric, which satisfy
\begin{equation}
  h_{\mu\nu} = \delta_{ab} e_\mu{}^a e_\nu{}^b,
  \qquad
  h^{\mu\nu} = \delta^{ab} e^\mu{}_a e^\nu{}_b,
\end{equation}
where $a=1,\ldots,d$ are flat frame indices.
The integration measure $e=\det(\tau_\mu,e_\mu{}^a)$ then is the determinant of these spatial vielbeine together with the clock one-form.

\subsection{Class of torsion-free metric compatible connections}
Consider the case of a symmetric connection that is metric compatible in the sense that
\begin{equation}
  \label{eq:nc-metric-compatibility}
  \nabla_\mu\tau_\nu=0\,,\qquad \nabla_\mu h^{\nu\rho}=0\,,
\end{equation}
such that $\Gamma^\rho_{[\mu\nu]}=0$. We can split such a connection as
\begin{equation}\label{eq:conredef}
    \Gamma^\rho_{\mu\nu}=\check\Gamma^\rho_{\mu\nu}+C^\rho{}_{\mu\nu}\,,
\end{equation}
where $C^\rho{}_{\mu\nu}$ is symmetric in $\mu$ and $\nu$ and transforms as a tensor under coordinate transformations, and where furthermore we defined 
\begin{equation}\label{eq:checkGamma}
    \check\Gamma^\rho_{\mu\nu}=-v^\rho\partial_\mu\tau_\nu+\frac{1}{2}h^{\rho\sigma}\left(\partial_\mu h_{\nu\sigma}+\partial_\nu h_{\mu\sigma}-\partial_\sigma h_{\mu\nu}\right)\,.
\end{equation}
Since we have assumed $\Gamma^\rho_{[\mu\nu]}=0$ in Trautman's postulate~\ref{it:traut3-metric-compatibility-symmetric} above, it follows that we have $d\tau=0$.
Demanding metric compatibility leads to
\begin{equation}
    C^\rho{}_{\mu\nu}\tau_\rho=0\,,\qquad C^\nu{}_{\mu\lambda}h^{\lambda\rho}+C^\rho{}_{\mu\lambda}h^{\nu\lambda}=0\,.
\end{equation}
We can solve the first of these conditions by writing $C^\rho{}_{\mu\nu}=h^{\rho\sigma}Y_{\sigma\mu\nu}$ where $Y_{\sigma\mu\nu}=Y_{\sigma\nu\mu}$. The second condition then tells us that
\begin{eqnarray}\label{eq:Ycondition}
\left(h^{\nu\sigma}h^{\lambda\rho}+h^{\rho\sigma}h^{\nu\lambda}\right)Y_{\sigma\mu\lambda}=0\,.
\end{eqnarray}
Using completeness on the $\mu$ index (i.e. invoking $\delta_\mu^\kappa=-\tau_\mu v^\kappa+h_{\mu\rho}h^{\rho\kappa}$ following Equation~\eqref{eq:first-nc-orthonormality} above), we can write
\begin{eqnarray}
Y_{\sigma\mu\nu}=-\frac{1}{2}\tau_\mu F_{\nu\sigma}-\frac{1}{2}\tau_\nu F_{\mu\sigma}+L_{\sigma\mu\nu}\,,
\end{eqnarray}
for some $F_{\mu\nu}$ and some $L_{\sigma\mu\nu}=L_{\sigma\nu\mu}$ which is purely spatial (meaning that all contractions with $v^\rho$ are zero). The factor of $-1/2$ is there for later convenience. From equation~\eqref{eq:Ycondition} we then get two equations by contracting with $v^\mu$ and with $h^{\mu\kappa}$,
\begin{eqnarray}
\left(h^{\nu\rho}h^{\lambda\sigma}+h^{\rho\sigma}h^{\nu\lambda}\right)F_{\nu\sigma}=0\,,\qquad L_{\sigma\mu\nu}=-L_{\nu\mu\sigma}\,.
\end{eqnarray}
It can be shown that $L_{\sigma\mu\nu}$ must be antisymmetric in its first two indices, since
\begin{equation}
    L_{\sigma\mu\nu}=L_{\sigma\nu\mu}=-L_{\mu\nu\sigma}=-L_{\mu\sigma\nu}\,.
\end{equation}
So we have a tensor $L_{\sigma\mu\nu}$ that is antisymmetric in slots 1 and 2 as well as in slots 1 and 3 and that is furthermore symmetric in slots 2 and 3.
Such a tensor is zero, as follows from
\begin{equation}
    L_{\sigma\mu\nu}=-L_{\mu\sigma\nu}=L_{\nu\sigma\mu}=L_{\nu\mu\sigma}\,,
\end{equation}
which shows that $L$ is symmetric in slots 1 and 3 but that means it must be zero since it is also antisymmetric in slots 1 and 3. Hence we have 
\begin{equation}
C^\rho{}_{\mu\nu}=-\frac{1}{2}h^{\rho\sigma}\left(\tau_\mu F_{\nu\sigma}+\tau_\nu F_{\mu\sigma}\right)\,,
\end{equation}
where $\left(h^{\nu\rho}h^{\lambda\sigma}+h^{\nu\lambda}h^{\rho\sigma}\right)F_{\nu\sigma}=0$ so that $F_{\mu\nu}$ is of the form
$F_{\mu\nu}=\tau_\mu X_\nu+\tau_\nu Y_\mu+X_{\mu\nu}$ where $X_{\mu\nu}$ is purely spatial and antisymmetric. Since only $h^{\rho\sigma}F_{\nu\sigma}$ appears in the expression for $C^\rho{}_{\mu\nu}$ we can without loss of generality set $Y_\mu=-X_\mu$ so that $F_{\mu\nu}$ is antisymmetric. 
Therefore we see that
\begin{equation}\label{eq:Gamma}
    \Gamma^\rho_{\mu\nu}=\check\Gamma^\rho_{\mu\nu}-\frac{1}{2}h^{\rho\sigma}\left(\tau_\mu F_{\nu\sigma}+\tau_\nu F_{\mu\sigma}\right)\,,
\end{equation}
where $\check\Gamma^\rho_{\mu\nu}$ is given in \eqref{eq:checkGamma}. Equation \eqref{eq:Gamma} is the most general torsion-free NC metric-compatible connection. Unlike in Lorentzian geometry, we see here that this connection is not unique, and its freedom is parametrized by the antisymmetric tensor $F_{\mu\nu}$~\cite{Dombrowski,Kuenzle1}.

We stress that $\check\Gamma^\rho_{\mu\nu}$ is not invariant under Galilean boosts, and therefore the covariant derivative whose connection coefficients are given by $\check\Gamma^\rho_{\mu\nu}$ does not form a proper affine connection (even though the covariant derivative transforms correctly under general coordinate transformations).
In order that $\Gamma^\rho_{\mu\nu}$ as defined in~\eqref{eq:Gamma} is boost-invariant, we must have that $F_{\mu\nu}$ transforms appropriately under Galilean boost transformations.
This is possible for a symmetric connection, and we will get back to this point further below.
With this caveat in mind, we will however still use a covariant derivative $\check\nabla_\mu$ with connection coefficients $\check\Gamma^\rho_{\mu\nu}$ and define an associated Riemann curvature $\check R_{\mu\nu\sigma}{}^\rho$ in the usual way. We refer to appendix \ref{sec:conventions} for our curvature tensor conventions.

\subsection{Ehlers conditions}
\label{ssec:ehlers}
For completeness, we briefly comment on the first Trautman condition, corresponding to the first equation of~\eqref{Trautman1}, even though we will not use it in the following.
First we show that it is equivalent to the Ehlers conditions, namely
\begin{eqnarray}
    &&1) \qquad R_{\mu\nu\sigma}{}^\rho R_{\lambda\kappa\rho}{}^\sigma h^{\nu\kappa}=0\,,\\
    && 2) \qquad R_{\mu\nu[\sigma}{}^\rho\tau_{\gamma]}=0\,,\\
    && 3)\qquad R_{\mu\nu\sigma}{}^{[\rho}h^{\gamma]\sigma}=0\,.
\end{eqnarray}
We will show that 1) is equivalent to 2) and that 2) is equivalent to 3). 

First we show that 1) implies 2). We project 1) with $v^\mu v^\lambda$.
In terms of tangent space indices this means that the condition becomes $R_{0abc}R_{0abc}=0$ where $R_{0abc}=-v^\mu e^\nu{}_a e^\sigma{}_b e_{\rho c}R_{\mu\nu\sigma}{}^\rho$
and where we used $R_{\mu\nu\sigma}{}^\rho\tau_\rho=0$ which follows from the covariant constancy of $\tau_\mu$.
This is a sum of squares and hence $R_{0abc}$.
Contracting condition 1) with $h^{\mu\lambda}$ we find $R_{abcd}R_{abcd}=0$ so that $R_{abcd}=0$.
From $R_{0bcd}=0$ and $R_{abcd}=0$ it follows that $R_{\mu\nu\sigma}{}^\rho h^{\sigma\gamma}=0$.
The latter equation is equivalent to condition 2).
This follows from $0=R_{\mu\nu\sigma}{}^\rho h^{\sigma\gamma}h_{\gamma\kappa}=R_{\mu\nu\kappa}{}^\rho+R_{\mu\nu\rho}{}^\sigma v^\rho\tau_\kappa$ where we used completeness. 

The other direction, that 2) implies 1), is more straightforward as 2) is equivalent to $R_{\mu\nu\sigma}{}^\rho h^{\sigma\gamma}=0$ which implies 1). Here we used again that $R_{\mu\nu\sigma}{}^\rho\tau_\rho=0$. To show that 2) is equivalent to 3) all we need is that $R_{\mu\nu\sigma}{}^{(\rho}h^{\gamma)\sigma}=0$ which follows from $[\nabla_\mu\,,\nabla_\nu]h^{\rho\gamma}=0$. This result implies that $R_{\mu\nu\sigma}{}^\rho h^{\sigma\gamma}=R_{\mu\nu\sigma}{}^{[\rho}h^{\gamma]\sigma}=0$ from which the result follows.

So what is the purpose of the first Trautman condition? To answer this we take a closer look at $R_{\mu\nu\sigma}{}^\rho h^{\sigma\gamma}=0$. Expressing this in terms of $\check R_{\mu\nu\sigma}{}^\rho$ and projecting by $h^{\mu\alpha}h^{\nu\beta}$ and $v^\nu h^{\mu\kappa}$ we find
\begin{eqnarray}
    0 & = & h^{\mu\alpha}h^{\nu\beta}\check R_{\mu\nu\sigma}{}^\rho h^{\sigma\gamma}\,,\\
    0 & = & h^{\rho\alpha}h^{\gamma\lambda}h^{\beta\kappa}\left(\check\nabla_\alpha K_{\lambda\beta}-\check\nabla_\lambda K_{\alpha\beta}+\frac{1}{2}\check\nabla_\beta F_{\alpha\lambda}\right)\,,
\end{eqnarray}
where $K_{\alpha\beta}=-\frac{1}{2}\mathcal{L}_v h_{\alpha\beta}$ with $\mathcal{L}_v$ the Lie derivative along $v^\mu$.
The first equation states that we must always have a flat geometry on constant time slices (these are the surfaces to which $\tau$ is the normal 1-form). The second equation is a condition on the magnetic part  of the field strength $F_{\mu\nu}$. However, these conditions are too strong. It should only be the equivalent of the Einstein equation that decides what the allowed spaces are.
In addition, the first Trautman condition does not follow from the $1/c$ expansion of Lorentzian geometry.

\subsection{Trautman condition}\label{subsec:Trautman}

The relation between the Riemann tensors associated with two connections related via \eqref{eq:conredef} is given by\,,
    \begin{equation}
    \label{eq:curvature-shift-in-connection}
        R_{\mu\nu\sigma}{}^\rho=\check R_{\mu\nu\sigma}{}^\rho-\check\nabla_\mu C^\rho{}_{\nu\sigma}+\check\nabla_\nu C^\rho{}_{\mu\sigma}-C^\rho{}_{\mu\lambda} C^\lambda{}_{\nu\sigma}+C^\rho{}_{\nu\lambda} C^\lambda{}_{\mu\sigma}\,,
    \end{equation}
    where on the left-hand side we have the Riemann tensor associated with $\nabla_\mu$ and on the right-hand side we have the Riemann tensor associated with $\check\nabla_\mu$. This gives for the Trautman condition \eqref{Trautman2} that
    \begin{equation}
       0= h^{\mu[\gamma}R_{\mu(\nu\sigma)}{}^{\rho]}= h^{\mu[\gamma}\check R_{\mu(\nu\sigma)}{}^{\rho]}+\frac{3}{4}h^{\mu\gamma}h^{\rho\kappa}\left(\tau_\nu\check\nabla_{[\mu}F_{\sigma\kappa]}+\tau_\sigma\check\nabla_{[\mu}F_{\nu\kappa]}\right)\,.
    \end{equation}
     It can be shown that $h^{\mu[\gamma}\check R_{\mu(\nu\sigma)}{}^{\rho]}=0$. The details can be found in appendix \ref{app:Trautman}. For now let us see what its consequences are. We find 
    \begin{equation}
       0= h^{\mu\gamma}h^{\rho\kappa}\left(\tau_\nu\check\nabla_{[\mu}F_{\sigma\kappa]}+\tau_\sigma\check\nabla_{[\mu}F_{\nu\kappa]}\right)\,.
    \end{equation}
    Contracting this equation with $v^\nu v^\sigma$ and with $v^\nu h^{\sigma\lambda}$ we see that this is true if and only if we have
    \begin{equation}\label{eq:BianchiF}
       0= \check\nabla_{[\mu}F_{\sigma\kappa]}= \partial_{[\mu}F_{\sigma\kappa]}\,,
    \end{equation}
    where we used that the torsion is zero.
    In other words the Trautman condition \eqref{Trautman2} leads to a Bianchi identity for $F_{\mu\nu}$.

    We point out that as an alternative to the Trautman condition \eqref{Trautman2} one could also impose
    \begin{equation}
    h_{\rho[\kappa}R_{\mu\nu]\sigma}{}^\rho v^\sigma=0\,,
    \end{equation}
    whenever $\Gamma^\rho_{\mu\nu}$ is symmetric. This is because of the following identity
        \begin{equation}
        h_{\rho[\kappa}R_{\mu\nu]\sigma}{}^\rho v^\sigma=-\nabla_{[\mu}F_{\nu\kappa]}=-\partial_{[\mu}F_{\nu\kappa]}\,,
    \end{equation}
    where we used \eqref{eq:Gamma}.

  From Equation~\eqref{eq:BianchiF} we thus see that the second Trautman condition~\eqref{Trautman2} is obeyed if we take $F_{\mu\nu}=\partial_\mu m_\nu-\partial_\nu m_\mu$ with $m_\mu$ defined up to a gauge transformations of the form $\delta m_\mu=\partial_\mu\sigma$.
  For the connection \eqref{eq:Gamma} to be invariant under Galilean boosts, we need $m_\mu$ to transform as $\delta m_\mu=\lambda_\mu$, where we remind the reader that $v^\mu\lambda_\mu=0$.
    Furthermore, in contrast to the first Trautman condition, the second Trautman condition does follow from the $1/c$ expansion of the following identity when working with a Lorentzian metric and the Levi-Civita connection, 
    \begin{equation}\label{eq:identity}
       g^{\mu[\gamma} R_{\mu(\nu\rho)}{}^{\sigma]}=0\,,
    \end{equation}
    where in this expression $R_{\mu\nu\rho}{}^{\sigma}$ is the curvature of the Levi-Civita connection. The $1/c$ expansion of \eqref{eq:identity}, assuming that $d\tau=0$, leads to \eqref{Trautman2}, which is equally identically satisfied in NC geometry since the $1/c$ expansion also tells us that $F_{\mu\nu}=\partial_\mu m_\nu-\partial_\nu m_\mu$ (see further below in Section~\ref{subsec:1/cGR}).

    \subsection{Field content of torsion-free Newton--Cartan geometry}

From now on we will always take $F_{\mu\nu}$ to be given by $\partial_\mu m_\nu-\partial_\nu m_\mu$. For this choice of $F_{\mu\nu}$ we will denote the connection coefficients by $\bar\Gamma^\rho_{\mu\nu}$, i.e. 
\begin{equation}\label{eq:gammabar}
    \bar\Gamma^\rho_{\mu\nu}=\check\Gamma^\rho_{\mu\nu}-\frac{1}{2}h^{\rho\sigma}\left(\tau_\mu F_{\nu\sigma}+\tau_\nu F_{\mu\sigma}\right)\,,
\end{equation}
where $F_{\mu\nu}=\partial_\mu m_\nu-\partial_\nu m_\mu$. The associated covariant derivatives and curvatures will be denoted by barred quantities.

The total field content of Newton--Cartan geometry is given by the gauge potential $m_\mu$
as well as $\tau_\mu$ and $h_{\mu\nu}$. These fields transform under the gauge transformations as
\begin{equation}\label{eq:gauge}
\delta\tau_\mu=0\,,\qquad\delta h_{\mu\nu}=\tau_\mu\lambda_\nu+\tau_\nu\lambda_\mu\,,\qquad\delta m_\mu=\partial_\mu\sigma+\lambda_\mu\,.
\end{equation}
where $v^\mu\lambda_\mu=0$ and where we left out the transformation under diffeomorphisms. The connection \eqref{eq:Gamma} is inert under these transformations. The symmetry of the connection implies that $d\tau=0$ and thus that time is absolute.

\subsection{Newton--Cartan gravity}

So far we have made no statements about dynamics. We now review how standard Newtonian gravity can be formulated in the framework of torsionless Newton--Cartan geometry.
Test particles will follow geodesics of the connection \eqref{eq:Gamma}. In other words, they are described by 
\begin{equation}\label{eq:NCgeodesic}
\ddot x^\mu+\bar\Gamma^\mu_{\nu\rho}\dot x^\nu\dot x^\rho=\ddot x^\mu+\check\Gamma^\mu_{\nu\rho}\dot x^\nu\dot x^\rho-h^{\mu\sigma}\dot x^\rho F_{\rho\sigma}=0\,,
\end{equation}
where the dots denote derivatives with respect to $\lambda$, the affine parameter along the geodesic. We have set $\tau_\mu \dot x^\mu=1$ which implies that the $\tau_\mu$ contraction of this equation is trivially satisfied. This geodesic equation follows from the action
\begin{equation}
\label{eq:geodesic} 
    S=m\int d\lambda\left[\frac{h_{\mu\nu}\dot x^\mu\dot x^\nu}{2\tau_\rho \dot x^\rho}-m_\mu\dot x^\mu\right]\,.
\end{equation}
This action has worldline reparametrisation invariance $\delta\lambda=\xi(\lambda)$ and $\delta x^\mu=\xi(\lambda)\dot x^\mu$ which can be used to fix $\tau_\mu \dot x^\mu=1$. The fact that $m$ appears as an overall parameter in this particle action, so that inertial and gravitational masses are equal, is a manifestation of the equivalence principle. The time-component of $m_\mu$ is the Newtonian potential.

Following Trautman's postulate~\ref{it:traut5-covariant-poisson}, the
equations corresponding to Newtonian gravity
should take the following covariant form 
in terms of the NC geometry,
\begin{equation}\label{eq:Newton}
  \bar R_{\mu\nu} = 8\pi G \frac{d-2}{d-1}\rho\tau_\mu\tau_\nu \spa d \tau =0 \, , 
\end{equation}
where $\bar R_{\mu\nu}=\bar R_{\mu\sigma\nu}{}^\sigma$ is the Ricci curvature associated with \eqref{eq:gammabar}, where $\rho$ is the mass density of a non-relativistic matter source, and where $(d+1)$ is the total spacetime dimension.
This corresponds to the Poisson equation of Newtonian
gravity written in an arbitrary frame. Equations \eqref{eq:Newton} follow from an action principle obtained from the $1/c^2$ expansion of the Einstein--Hilbert action coupled to the action of a massive point particle \cite{Hansen:2020pqs}, as we will review in Section~\ref{ssec:type-I-tnc-mass-coupling}.

Rewriting the equations for NC gravity~\eqref{eq:Newton}, which are expressed in terms of the $\bar\Gamma^\rho_{\mu\nu}$ connection to a set of equations that are expressed in terms of the $\check\Gamma^\rho_{\mu\nu}$ connection, the equations for NC gravity can then be written as (where we have contracted \eqref{eq:Newton} with $v^\mu$ and $h^{\mu\nu}$),
\begin{eqnarray}
h^{\mu\rho}h^{\nu\sigma}\check R_{\rho\sigma} & = & 0\,,\label{eq:Ricciflat}\\
h^{\mu\rho}v^{\sigma}\check R_{\rho\sigma} & = & -\frac{1}{2}e^{-1}\partial_\lambda\left(eh^{\mu\rho}h^{\lambda\sigma}F_{\rho\sigma}\right)\,,\label{eq:NCG2}\\
v^{\rho}v^{\sigma}\check R_{\rho\sigma} & = & -e^{-1}\partial_\rho\left(e v^\nu h^{\rho\sigma}F_{\nu\sigma}\right)-\frac{1}{4}h^{\mu\nu}h^{\rho\sigma}F_{\mu\rho}F_{\nu\sigma}+8\pi G \frac{d-2}{d-1}\rho\,.\label{eq:NCG3}
\end{eqnarray}
In NC gravity this should be supplemented with the condition $d \tau=0$. 
The left hand side is pure geometric data and the right hand side depends entirely on the ``electric'' and ``magnetic'' field strength components of $F_{\mu\nu}$. We will see below that the divergence of the electric field strength in the third equation, i.e. $e^{-1}\partial_\rho\left(e v^\nu h^{\rho\sigma}F_{\nu\sigma}\right)$, is what gives rise to Newton's law of gravity. In order to arrive at Newtonian gravity we need to somehow get rid of the magnetic field strength term, $h^{\mu\nu}h^{\rho\sigma}F_{\mu\rho}F_{\nu\sigma}$, in Equation~\eqref{eq:NCG3}. This is the rationale behind the first Trautman condition. However, it was later realized that this condition is not necessary as one can argue that the magnetic field strength has to be zero as a result of a boundary condition that states that 
$h^{\mu\nu}h^{\rho\sigma}F_{\mu\rho}$ has to vanish at infinity. This latter condition follows from the $1/c$ expansion of a metric that is asymptotically flat, and this is therefore a more minimal approach to recovering Newtonian gravity.

\subsection{Gauge fixing}
To recover Newtonian gravity in its usual form, it is unavoidable to talk about gauge fixing the Newton--Cartan gauge symmetries \eqref{eq:gauge} and diffeomorphisms. A covariant definition of a locally flat Newton--Cartan geometry could be to simply require $\bar R_{\mu\nu\sigma}{}^\rho=0$. This condition is invariant under the Galilean boost and $\sigma$-gauge transformations of equation~\eqref{eq:gauge}.
However, in practice, only $\tau_\mu$ and $h_{\mu\nu}$ are often treated as geometric fields, whereas $m_\mu$ is interpreted as a force field. This is not a covariant distinction and is largely the result of historical bias. For example, one can always use \eqref{eq:gauge} to gauge away $m_\mu$ entirely. This has been shown in \cite{Bekaert:2014bwa}, see \cite{Kapustin:2021omc} for some examples.

A common gauge fixing that can always be made when $d\tau=0$ is the following. First, we partially fix diffeomorphisms so that $\tau=dt$ where $t$ is our time coordinate. In this class of coordinate systems, the boosts act as $\delta h_{ti}=\lambda_i$, where the latter is an arbitrary one-form in $d$ spatial dimensions. We can thus completely fix the Galilean boost symmetry by demanding that $h_{ti}=0$ in these coordinates. We then find from $v^\mu h_{\mu\nu}=0$ and $v^\mu\tau_\mu=-1$ that $v^t=-1$ and $h_{tt}=0$ and furthermore that $v^i=0$, where we used that $h_{ij}$ is invertible with inverse $h^{ij}$. Using $\tau_\mu h^{\mu\nu}=0$ we find $h^{tt}=0=h^{ti}$. To summarise, we can always go to a gauge in which 
\begin{equation}\label{eq:gaugefixing}
  \tau=dt\,,\qquad h_{\mu\nu}dx^\mu dx^\nu=h_{ij} dx^i dx^j\,,\qquad v=-\partial_t\,,\qquad h^{\mu\nu}\partial_\mu\partial_\nu=h^{ij}\partial_i\partial_j\,,
\end{equation}
and where $m_\mu$ is completely arbitrary. This gauge choice still has residual gauge transformations acting on it.
Demanding that the geometry on constant time slices is flat leads to the condition that $h^{\alpha\mu}h^{\beta\nu}h^{\gamma\sigma}\check R_{\mu\nu\sigma}{}^\rho=0$ (which is invariant under Galilean boosts).
In our partially fixed gauge \eqref{eq:gaugefixing} this amounts to demanding that the Riemann tensor of the Riemannian metric $h_{ij}$ is zero.
We can fix diffeomorphisms further so that locally $h_{ij}=a^2(t)\delta_{ij}$ for some function $a(t)$.
This function plays an important role in Newtonian cosmology.
We can perform a coordinate transformation of the form $x'^i=a(t)x^i$ and $t'=t$ followed by a finite Galilean boost so that in the primed coordinate system $h'_{\mu\nu}dx'^\mu dx'^\nu=\delta_{ij}dx'^i dx'^j$ and $\tau=dt'$. This transformation will of course affect the $m_\mu$ connection but the point here is to fix $\tau_\mu$ and $h_{\mu\nu}$ as much as possible. We can thus without loss of generality set $a(t)=1$. Hence we will continue by working with
\begin{equation}
  \tau=dt\,,\qquad h_{\mu\nu}dx^\mu dx^\nu=\delta_{ij} dx^i dx^j\,,\qquad v=-\partial_t\,,\qquad h^{\mu\nu}\partial_\mu\partial_\nu=\delta^{ij}\partial_i\partial_j\,.
\end{equation}

In this case all the nontrivial information about the NC geometry is in $m_\mu$, which at this point is still fully arbitrary.
In this gauge and with these choices the equations \eqref{eq:Ricciflat}--\eqref{eq:NCG3} reduce to
\begin{eqnarray}
0 & = & \partial_i F_{ij}\,,\\
0 & = & \partial_i F_{ti}-\frac{1}{4}F_{ij}F_{ij}+8\pi G \frac{d-2}{d-1}\rho\,.\label{eq:intermediateNewton}
\end{eqnarray}
The first of these two equations is solved by $F_{ij}=\varepsilon_{ijk}\partial_k F$. The Bianchi identity $\partial_{[i}F_{jk]}=0$ (which follows from the Trautman condition) then tells us that $F$ is harmonic on flat space (without any sources). By Liouville's theorem this function must be constant in order to be regular everywhere (including infinity) and hence we conclude that $F_{ij}=0$. We can gauge fix $m_i=0$, using the freedom to transform $m_\mu$ as $\delta m_\mu=\partial_\mu\sigma$. Now the NC gravity equations simplify to the well-known Poisson equation
\begin{equation}
\nabla^2 \Phi_{\rm N} = 8\pi G \frac{d-2}{d-1}\rho\,,
\end{equation}
where we defined $m_t=\Phi_N$, the Newtonian potential.
Therefore we see that the covariant equation~\eqref{eq:Newton} indeed reproduces the usual form of Newtonian gravity in an appropriate coordinate system.

\subsection{Large speed of light expansion of GR}\label{subsec:1/cGR}

We very briefly review how Trautman's definition of NC gravity follows from the $1/c$ expansion of GR. We will have much more to say about the $1/c$ expansion in Section~\ref{sec:NRexpGR} so we will keep it brief. The following is essentially Dautcourt's work \cite{Dautcourt1,Dautcourt2}. We will only consider even powers of $1/c^2$ and assume analyticity in $1/c^2$. The metric expands as
\begin{equation}
    g_{\mu\nu}=-c^2\tau_\mu\tau_\nu+h_{\mu\nu}-\tau_\mu m_\nu-\tau_\nu m_\mu+\mathcal{O}(c^{-2})\,,
\end{equation}
where $h_{\mu\nu}$ has signature $(0,1,\ldots,1)$ and where $\tau_\mu\tau_\nu+h_{\mu\nu}$ is invertible. The Christoffel connection expands as
\begin{eqnarray}
    \Gamma^\rho_{\mu\nu}=\bar \Gamma^\rho_{\mu\nu}+\mathcal{O}(c^{-2})\,,
\end{eqnarray}
where $\bar\Gamma^\rho_{\mu\nu}$ is given in Equation~\eqref{eq:gammabar}, but only provided we set $d\tau=0$ by hand. If we also expand the diffeomorphisms parameter $\Xi^\mu=\xi^\mu+c^{-2}\zeta^\mu+\mathcal{O}(c^{-4})$ then $m_\mu$ transforms as $\delta m_\mu=\partial_\mu\sigma$ with $\sigma=\tau_\mu\zeta^\mu$ under the subleading diffeomorphisms with parameter $\zeta^\mu$, again only provided that $d\tau=0$. Finally, the combination $h_{\mu\nu}-\tau_\mu m_\nu-\tau_\nu m_\mu$ is invariant under the Galilean boost transformation with parameter $\lambda_\mu$ discussed earlier.
We see that we recover the NC fields and their gauge properties as discussed above, but only under the assumption that $d\tau=0$.
The $1/c$ expansion of matter will be discussed in later sections, but an important observation is that one can only obtain the NC geodesic equation \eqref{eq:NCgeodesic} from a $1/c$ expansion if $d\tau=0$. The $1/c$ expansion of the Einstein equations coupled to the energy-momentum tensor of a massive point particle then leads to \eqref{eq:Newton}. If $g_{\mu\nu}$ is asymptotically flat it follows that $m_\mu$ (in the limit $r\rightarrow\infty$) is at most pure gauge at spatial infinity. This is why, below \eqref{eq:intermediateNewton}, we used that $F_{ij}$ is non-singular at infinity.

\section{Recent history: revival and new developments}\label{sec:recenthistory}

The last decade has seen a surge of interest in the topic of non-Lorentzian geometries, and in particular of Newton--Cartan (NC) geometry. To separate this new development from the older work reviewed above we treat the paper \cite{Andringa:2010it} as the beginning of this new development. This work has played an important role in later developments. It showed that for $d\tau=0$ it is possible to view Newton--Cartan gravity as the dynamics of a geometry that can be obtained by gauging the Bargmann algebra subject to appropriate curvature constraints. We will review this approach further below. This approach made conditions such as the Trautman condition \eqref{Trautman2} obsolete as the latter now follows trivially from a Bianchi identity associated with the field strength that appears in the gauging procedure.

Another major step forward was the paper \cite{VandenBleeken:2017rij} where it was realized that the $1/c$ expansion of GR can be done in full generality without imposing by hand (or via boundary conditions and assumptions about sources) the requirement that $d\tau=0$. This led to the realization that a non-relativistic approximation can describe gravitational fields that are e.g. strong on the scale of the Schwarzschild radius. 

In textbook non-relativistic approximations of GR, one also assumes that one is in a weak-field regime. In this case, GR time dilation effects are (in part) described by the Newtonian potential. In the case of a strong field non-relativistic regime, GR time dilation is (in part) described by a NC lapse function $N$ such that $\tau=NdT$ for scalars $N$ and $T$. In other words, gravitational time dilation could be incorporated into the framework of NC geometry by allowing for $d\tau\neq 0$, which turns out to correspond to nonzero torsion in the Newton--Cartan connection.
Not all torsion is allowed, and at least on shell, we still need to impose the requirement that $\tau\wedge d\tau=0$, which guarantees that spacetime can be consistently decomposed into spatial submanifolds.

It seemed natural that in order to describe NC geometry with $d\tau\neq 0$ all one had to do was to extend the gauging methods of \cite{Andringa:2010it} to this more general case with nonzero torsion. However, it turned out that is not the right thing to do.
One cannot consistently couple Newton--Cartan geometry with local Bargmann symmetry to matter sources without turning on torsion, which is incompatible with the usual description of Newtonian gravity as discussed in Section~\ref{sec:nc-basics}.
Furthermore, the resulting torsion even violates the condition $\tau\wedge d\tau=0$, as we demonstrate explicitly in Section~\ref{ssec:type-I-tnc-mass-coupling} below.
Instead, the relevant algebra is a different one, and its gauging leads to the correct extension of NC geometry when the torsion is nonzero.
To distinguish this new framework of torsional NC geometry from the one obtained by gauging the Bargmann algebra we refer to the latter as type I (gauging of Bargmann) and to the former as type II torsional NC geometry ($1/c^2$ expansion). 
These notions of geometry coincide if the torsion vanishes.

We now focus on reviewing type I torsional Newton--Cartan geometry, its relation to the gauging of the Bargmann algebra and its gravitational action in Section~\ref{sec:type-I-tnc}.
Once we have seen the incompatibility of the latter with the standard formulation of Newtonian gravity, we introduce type II torsional Newton--Cartan geometry in Section~\ref{sec:NRexpGR}.

\section{Type I torsional Newton-Cartan geometry}
\label{sec:type-I-tnc}
As is well known, one can obtain Lorentzian geometry through a gauging of the Poincaré algebra.
After quickly reviewing this construction, we show how one can obtain type~I torsional Newton--Cartan (TNC) geometry from a gauging of the Bargmann algebra.
We subsequently discuss how a $(d+1)$-dimensional TNC geometry can also be obtained from the following two constructions:
\begin{itemize}
\item a null reduction of a $(d+2)$-dimensional Lorentzian geometry,
\item a large speed of light limit of a $(d+1)$-dimensional Lorentzian geometry with an electromagnetic background field.
\end{itemize} 
Next, we show how the action equivalent of these constructions allows one to find the action of a non-relativistic particle probe as well as the dynamics of type~I TNC spacetime itself.
We then demonstrate that the resulting gravity actions for dynamical type~I TNC geometry lead to nonzero torsion in the presence of mass sources.
In this sense, type~I TNC geometry is not appropriate to describe the zero-torsion limit corresponding to Newtonian gravity in the presence of matter.
Finally, we show that the same arguments still apply if one only works on shell by null reducing Einstein's equations.

\subsection{The Bargmann algebra}
\label{ssec:type-I-tnc-bargmann-alg}
As we will review momentarily, one can obtain Lorentzian geometry by applying a gauging procedure to the Poincaré algebra.
Similarly, type I Newton--Cartan geometry can be obtained from a gauging of the Bargmann algebra, which encodes its local symmetries.
In this section, we first show how the Bargmann algebra can be obtained from a contraction and from a null reduction of the Poincaré algebra.
These two derivations will be mirrored at the level of the point particle and gravity actions later on.

First, we consider an İnönü--Wigner contraction of the Poincaré algebra trivially extended with a $U(1)$ generator that commutes with the entire Poincar\'e algebra, which we denote by $Q$.
As we will see in the following, this generator is associated to an electromagnetic coupling.
We work in $(d+1)$ spacetime dimensions and use the following conventions for the Poincaré algebra,
\begin{subequations}
  \label{eq:poincare-commutation-relations}
  \begin{align}
    [M_{AB}, M_{CD}]
    &= \eta_{AC} M_{BD}
    - \eta_{BC} M_{AD}
    + \eta_{BD} M_{AC}
    - \eta_{AD} M_{BC}\,,
    \\
    [M_{AB}, P_C]
    &= \eta_{AC} P_B - \eta_{BC} P_A\,.
  \end{align}
\end{subequations}
Here, $P_A$ are translation generators
and $M_{AB}=-M_{AB}$ are the Lorentz and rotation generators.
Now we consider a space-time split of the $(d+1)$-dimensional algebra indices $A=(0,a)$, where $a=1,\cdots,d$ is a spatial index, and we define the generators
\begin{equation}
  \label{eq:bargmann-contraction-defs}
  H = c P_0 + Q\,,
  \quad
  N = \frac{1}{c} P_0\,,
  \quad
  G_a = \frac{1}{c} M_{0a}\,,
  \quad
  J_{ab} = M_{ab}\,.
\end{equation}
So far, this is just a change of basis.
However, if we now take the limit $c\to\infty$, we end up with an inequivalent algebra whose nonzero commutation relations are
\begin{subequations}
  \label{eq:bargmann}
  \begin{align}
    [J_{ab}, J_{cd}]
    &= \delta_{ac} J_{bd}
    - \delta_{bc} J_{ad}
    + \delta_{bd} J_{ac}
    - \delta_{ad} J_{bc}\,,
    \\
    [J_{ab}, P_c]
    &= \delta_{ac} P_b - \delta_{bc} P_a\,,
    &&&
    [J_{ab}, G_c]
    &= \delta_{ac} G_b - \delta_{bc} G_a\,,
    \\
    [G_a,H]
    &= - P_a\,,
    &&&
    [G_a,P_b]
    &= - \delta_{ab} N\,.
  \end{align}
\end{subequations}
This is the Bargmann algebra.
For $N=0$ we recover the Galilei algebra, which contains time translations $H$, spatial translations $P_a$ and rotations $J_{ab}$ as well as the Galilean boosts $G_a$. For $N\neq 0$ we obtain the centrally extended Galilei algebra known as the  Bargmann algebra. The central extension $N$, which, as we will see later, corresponds to the gauge potential $m_\mu$ that was introduced previously.

Another way to obtain the algebra~\eqref{eq:bargmann} is from a null reduction of
the Poincaré algebra in $(d+2)$ dimensions~\cite{Gomis:1978mv}.
If we consider all generators of this algebra that commute with the null translation generator
$N = P_+ = (P_0 + P_{d+1})/\sqrt{2}$,
we get
\begin{equation}
  \begin{gathered}
    \label{eq:null-centralizer-generators}
    P_a\,,
    \qquad
    H = P_-
    = (P_0-P_{d+1})/\sqrt{2}\,,
    \\
    J_{ab} = M_{ab}\,,
    \qquad
    G_a = M_{+a}
    = (M_{0a} + M_{(d+1)a})/\sqrt{2}\,,
  \end{gathered}
\end{equation}
which form a subalgebra corresponding to the Bargmann algebra~\eqref{eq:bargmann}.
Note that the higher-dimensional Lorentz boosts $M_{-a}$ and $M_{+-}$ do not commute with $H$ and therefore do not enter in this subalgebra.

\subsection{Type I torsional Newton--Cartan geometry}
\label{ssec:type-I-tnc-geometry}
As was first realized in~\cite{Andringa:2010it}, type I Newton--Cartan geometry can be obtained by gauging the Bargmann algebra~\eqref{eq:bargmann}.
We review this gauging construction in this section, including its generalization to non-zero torsion~\cite{Hartong:2015zia}.%
\footnote{%
  Type I Newton--Cartan geometry with torsion can also be derived from the Noether procedure~\cite{Festuccia:2016awg}.
}
Before that, we briefly review the well-known procedure for obtaining Lorentzian geometry from a gauging of the Poincaré algebra.
Finally, we also show how type I Newton--Cartan geometry can be obtained from a null reduction of a higher-dimensional Lorentzian geometry.

\subsubsection{Gauging Poincaré}
\label{sssec:type-I-tnc-geometry-gauging-poincare}
The gauging procedure starts from a connection valued in the algebra in question,%
\footnote{%
  See also the recent review~\cite{Bergshoeff:2022eog} for more background on non-Lorentzian geometries in general, as well as~\cite{Figueroa-OFarrill:2022mcy} for a more mathematical perspective on the gauging procedure.
}
\begin{equation}
  \label{eq:gauging-poincare-connection-decomposition}
  \mathcal{A}_\mu
  = E_\mu{}^A P_A + \frac{1}{2} \Omega_\mu{}^{AB} M_{AB}\,,
\end{equation}
where $A,B=0,1,\ldots, d$ and whose coefficients $E_\mu{}^A$ and $\Omega_\mu{}^A{}_B$ can be interpreted as frame fields or vielbeine and a `spin' connection for the frames.
Indices are raised and lowered using the Minkowski metric $\eta_{AB}$ on the frame bundle.
The curvature of the total connection $\mathcal{A}$ is then
\begin{align}
  \label{eq:gauge-connection-curvature}
  \mathcal{F}_{\mu\nu}
  &=  \partial_\mu\mathcal{A}_\nu-\partial_\nu\mathcal{A}_\mu+[{\mathcal{A}}_{\mu}\,,{\mathcal{A}}_{\nu}]\nonumber\\
  &= R(P)_{\mu\nu}{}^A P_A
  + \frac{1}{2} R(M)_{\mu\nu}{}^{AB} M_{AB}\,,
\end{align}
whose components correspond to the torsion and the curvature of the frame connection.
The connection $\mathcal{A}_\mu$ transforms in the adjoint of the Poincaré algebra,
\begin{equation}
  \label{eq:gauge-connection-transformation}
  \delta{\mathcal{A}}_{\mu}=\partial_\mu\Lambda+[{\mathcal{A}}_{\mu}\,,\Lambda]\,,
\end{equation}
where the transformation parameter $\Lambda$ can likewise be decomposed in $P_A$ and $M_{AB}$ components.
However, given a vector field $\xi^\mu$ on the base manifold, it is useful to parametrize $\Lambda$ instead using
\begin{equation}
  \label{eq:gauging-poincare-shifted-param}
  \Lambda=\xi^\mu\mathcal{A}_\mu + \Sigma\,,
  \qquad
  \Sigma = \frac{1}{2} \Lambda^{AB} M_{AB}\,,
\end{equation}
which can be done without loss of generality.
We can then define a distinct transformation $\bar\delta$ acting on $\mathcal{A}_\mu$ as follows,
\begin{equation}
  \label{eq:bardeltaA-def}
  \bar\delta\mathcal{A}_\mu
  =\delta\mathcal{A}_\mu-\xi^\nu \mathcal{F}_{\mu\nu}
  =\mathcal{L}_\xi\mathcal{A}_\mu+\partial_\mu\Sigma+[{\mathcal{A}}_{\mu}\,,\Sigma]\,.
\end{equation}
At this point,
we recover the Lie derivative $\mathcal{L}_\xi$ along $\xi^\mu$,
and we have effectively exchanged the gauge transformations along $P_A$ for diffeomorphisms\footnote{Strictly speaking, once we work with the $\bar\delta$ transformations we are no longer gauging the Poincar\'e algebra, but we are passing to the Cartan geometry modelled on the Klein pair consisting of the Poincar\'e algebra and the Lorentz algebra generated by $M_{AB}$.
See \cite{Figueroa-OFarrill:2022mcy} for more details.}.

In works on gauging space-time symmetry groups, it is often suggested that diffeomorphisms can only be obtained once specific curvature constraints are imposed.%
\footnote{%
  This is because setting to zero some of the curvatures in $\mathcal{F}_{\mu\nu}$ identifies $\bar\delta$ with $\delta$ in \eqref{eq:bardeltaA-def} for those fields that are not fixed by the curvature constraints.
  There is no need for the $\delta$ and $\bar\delta$ transformations to coincide.
}
We emphasize that the transformation $\bar\delta\mathcal{A}_\mu$ can be considered for any value of the total curvature $\mathcal{F}_{\mu\nu}$, including nonzero torsion $R(P)_{\mu\nu}{}^A$.
While this extension is often not of immediate interest in Lorentzian geometry, it is crucial in non-Lorentzian geometry.

Expanding~\eqref{eq:bardeltaA-def} in components, we obtain the transformation rules
\begin{subequations}
  \begin{align}
    \label{eq:gauging-poincare-vielbein-deltabar-transformations}
    \bar\delta E_\mu{}^A
    &= \LL_\xi E_\mu{}^A + \Lambda^A{}_B E_\mu{}^B\,,
    \\
    \bar\delta \Omega_\mu{}^A{}_B
    &= \LL_\xi \Omega_\mu{}^A{}_B
    + \pd_\mu \Lambda^A{}_B
    + \Lambda^A{}_C \Omega_\mu{}^C{}_B
    - \Lambda^C{}_B \Omega_\mu{}^A{}_C\,,
  \end{align}
\end{subequations}
corresponding to diffeomorphisms and local Lorentz transformations.
Next, we can introduce a set of inverse vielbeine $E^\mu{}_A$ such that
\begin{equation}
  \label{eq:gauging-poincare-inverse-vielbeine}
  E_\mu{}^A E^\mu{}_B
  = \delta^A_B\,,
  \qquad
  E_\mu{}^A E^\nu{}_A
  = \delta_\mu^\nu\,.
\end{equation}
Using these properties, the transformations of the vielbeine~\eqref{eq:gauging-poincare-vielbein-deltabar-transformations} imply
\begin{equation}
  \bar\delta E^\mu{}_A
  = \LL_\xi E^\mu{}_A - \Lambda^B{}_A E^\mu{}_B\,.
\end{equation}
Then we can construct the Lorentzian metric and its inverse,
\begin{equation}
  \label{eq:gauging-poincare-lorentzian-metric}
  g_{\mu\nu}=\eta_{AB} E_\mu{}^A E_\nu{}^B\,,
  \qquad
  g^{\mu\nu}=\eta^{AB} E^\mu{}_A E^\nu{}_B\,,
\end{equation}
which are invariant under local Lorentz transformations.

Finally, note that
we can translate between the connection $\Omega_\mu{}^A{}_B$ in the frame bundle and the affine connection $\Gamma^\rho_{\mu\nu}$ using the vielbein postulate
\begin{equation}
  \pd_\mu E_\nu{}^A
  + \Omega_{\mu}{}^A{}_B E_\nu{}^B
  - \Gamma^\rho_{\mu\nu} E_\rho{}^A
  = 0\,.
\end{equation}
Since $\Omega_\mu{}^{AB}$ is antisymmetric in its frame indices $AB$ due to its definition in~\eqref{eq:gauging-poincare-connection-decomposition},
the covariant derivative~$\nabla_\mu$ of the corresponding affine connection is always compatible with the Lorentzian metric~\eqref{eq:gauging-poincare-lorentzian-metric}.
However, since the torsion
$2\Gamma^\rho_{[\mu\nu]} = E^\rho{}_A R(P)_{\mu\nu}{}^A$
is not necessarily zero, this connection is not necessarily equal to the usual Levi-Civita connection.
In the following, we will mainly work directly with affine connections instead of the frame bundle connections.
However, for the metric variables, the gauging procedure outlined above is a useful method for obtaining the transformations of geometric quantities under local symmetries, even in the presence of torsion.

\subsubsection{Gauging Bargmann}
\label{sssec:type-I-tnc-geometry-gauging-bargmann}
Now we repeat the gauging procedure for the Bargmann algebra~\eqref{eq:bargmann}, and we show that this leads to type I Newton--Cartan geometry with torsion.
Our starting point is now
\begin{equation}
  \label{eq:gauging-bargmann-connection-decomposition}
  \mathcal{A}_\mu
  = H\tau_{\mu} + e_\mu{}^aP_a
  + G_a\Omega_{\mu}{}^a
  + \frac{1}{2}\Omega_{\mu}{}^{ab} J_{ab}
  + N m_\mu\,,
\end{equation}
where we use $\delta_{ab}$ to raise and lower spatial indices and where $\tau_\mu$ is the clock one-form,
the $e{}_\mu^a$ are spatial vielbeine,
$\Omega_\mu{}^a$ and $\Omega_\mu{}^a{}_b$ are frame connections
and $m_\mu$ is the Bargmann gauge potential associated with the central element $N$.
Following~\eqref{eq:gauging-poincare-shifted-param}, we parametrize the total gauge parameter as
\begin{equation}
  \Lambda=\xi^\mu\mathcal{A}_\mu+\Sigma\,,
  \qquad
  \Sigma = G_a\lambda^a+\frac{1}{2}J_{ab}\lambda^{ab} + N\sigma\,,
\end{equation}
Then the $\bar\delta$ transformations defined in~\eqref{eq:bardeltaA-def} lead to
\begin{subequations}
  \label{eq:deltabar-bargman-trafos}
  \begin{align}
    \bar\delta\tau_\mu
    &= \mathcal{L}_\xi\tau_\mu\,,
    \label{eq:trafo1}
    \\
    \bar\delta e_\mu{}^a
    &= \mathcal{L}_\xi e_\mu{}^a+\lambda^a{}_b e_\mu{}^b
    +\lambda^a\tau_\mu\,,\label{eq:trafo2}
    \\
    \bar\delta\Omega_\mu{}^a
    &= \mathcal{L}_\xi\Omega_\mu{}^a+\partial_\mu\lambda^a+\lambda^a{}_b \Omega_\mu{}^b+\lambda^b\Omega_{\mu b}{}^a\,,\label{eq:trafo3}
    \\
    \bar\delta\Omega_\mu{}^{ab}
    &= \mathcal{L}_\xi\Omega_\mu{}^{ab}+\partial_\mu\lambda^{ab}+2\lambda^{[a}{}_c\Omega_\mu{}^{|c|b]}\,,
    \\ 
    \label{eq:trafom}
    \bar\delta m_\mu
    &= \mathcal{L}_\xi m_\mu+\partial_\mu\sigma+e_\mu{}^a\lambda_a\,,
  \end{align}
\end{subequations}
where $\lambda^a$, $\lambda^{ab}$ and $\sigma $ are the parameters of the local Galilean boosts, local rotations and local $U(1)$ Bargmann transformations, respectively.
Likewise, following~\eqref{eq:gauging-poincare-inverse-vielbeine}, we can define a set of inverse vielbeine $(\tau^\mu,e_\mu{}^a)$ that satisfy
\begin{equation}
  v^\mu\tau_\mu=-1\,,
  \qquad
  v^\mu e_\mu{}^a=0\,,
  \qquad
  e^\mu{}_a\tau_\mu=0\,,
  \qquad
  e^\mu{}_a e_\mu{}^b=\delta^b_a\,.
\end{equation}
Using~\eqref{eq:trafo1} and~\eqref{eq:trafo2}, their transformations are
\begin{equation}
  \bar\delta v^\mu
  = \LL_\xi v^\mu - \lambda^a e^\mu{}_a\,,
  \qquad
  \bar\delta e^\mu{}_a
  = \LL_\xi e^\mu{}_a - \lambda^b{}_a e^\mu_b\,.
\end{equation}
Based on this, we can define the rotation-invariant spatial tensors as
\begin{equation}
  \label{eq:type1-tnc-spatial-metric-and-inverse}
  h_{\mu \nu} = \delta_{ab} e_\mu^a e_\nu^b\,,
  \qquad
  h^{\mu \nu} = \delta^{ab} e^\mu_a e^\nu_b\,,
\end{equation}
which satisfy the orthonormality relations
\begin{equation}
  v^\mu\tau_\mu=-1\,,
  \qquad
  v^\mu h_{\mu\nu}=0\,,
  \qquad
  h^{\mu\nu}\tau_\nu=0\,,
  \qquad
  h^{\mu\rho} h_{\rho\nu}
  = \delta^\mu_\nu + v^\mu \tau_\nu\,.
\end{equation}
With this, we have obtained the full field content of type I Newton--Cartan geometry, which consists of
a timelike one-form $\tau_\mu$, a spatial symmetric tensor $h_{\mu \nu} $ of signature $(0,1,\ldots,1)$ and a $U(1)$ gauge field $m_\mu$ associated with the central Bargmann mass generator $N$.
All these objects are spacetime tensors, since they transform by a Lie derivative under diffeomorphisms~$\xi^\mu$.
In addition, they transform as
\begin{equation}
  \label{eq:type-I-tnc-metric-boost-tr}
  \bar\delta\tau_\mu =0\,,
  \qquad
  \bar \delta h_{\mu \nu} = \lambda_\mu \tau_\nu + \lambda_\nu \tau_\mu\,,
  \qquad
  \bar\delta m_\mu  = \partial_\mu\sigma+ \lambda_\mu \,, 
\end{equation}
under local Galilean boosts $\lambda_\mu  =  e^a_\mu \lambda^a$ and $U(1)$ gauge transformations $\sigma$.
Note that we thus recover the transformations \eqref{eq:gauge} given before.  
Similarly, the transformations of the inverse timelike vielbein and spatial co-metric are
\begin{equation}
  \bar\delta v^\mu =  h^{\mu \nu} \lambda_\nu\,,
  \qquad
  \bar \delta h^{\mu \nu} = 0 \,.
\end{equation}
We thus see that both $\tau_\mu$ and $h^{\mu \nu}$ are invariant under Galilean boost.
Additionally, using the $U(1)$ gauge field $m_\mu$, we can construct the following boost-invariant combinations,
\begin{subequations}
  \label{eq:boost-invar-tnc-variables}
  \begin{align}
    \hat v^\mu
    &= v^\mu-h^{\mu\nu}m_\nu\,,
    \label{eq:hatv}
    \\ 
    \bar h_{\mu\nu}
    &= h_{\mu\nu}-\tau_\mu m_\nu-\tau_\nu m_\mu\,,
    \label{eq:barh}
    \\
    \hat\Phi
    &= -v^\mu m_\mu+\tfrac{1}{2}h^{\mu\nu}m_\mu m_\nu\,.
    \label{eq:hatPhi}
  \end{align}
\end{subequations}
Given an appropriate gauge choice, we will see later that the first term in $\hat\Phi$ essentially plays the role of the Newtonian gravitational potential.

\subsubsection{Affine connection, torsion and curvature}
\label{sssec:type-I-tnc-geometry-connection-curvature}
Given the metric variables for type I Newton--Cartan geometry and their transformations that we obtained from the gauging procedure above, we would like to construct the closest analogue of the Levi-Civita connection for Newton--Cartan geometry.%
\footnote{We refer the reader to Sections 2-7 in \cite{Hartong:2015zia} for further details relevant to this subsection.} 
Our starting point is the TNC analogue of metric compatibility
\begin{equation}\label{eq:conditionGamma1}
  \nabla_\mu\tau_\nu=0\,,
  \qquad
  \nabla_\mu h^{\nu\rho}=0\,.
\end{equation}
These conditions do not uniquely specify the connection in terms of the geometry.
Following Equation~\eqref{eq:conredef} above, one can show that the general solution takes the form
\begin{align}
  \label{eq:generalformY}
  \Gamma^{\rho}_{\mu\nu}
  &= \check\Gamma^\rho_{\mu\nu}
  + C^\rho{}_{\mu\nu}\,,
  \\
  \label{eq:gamma-check-tnc}
  \check\Gamma^\rho_{\mu\nu}
  &= -v^\rho\partial_\mu\tau_\nu+\frac{1}{2}h^{\rho\sigma}\left(\partial_\mu h_{\nu\sigma}+\partial_\nu h_{\mu\sigma}-\partial_\sigma h_{\mu\nu}\right)\,,
  \\
  C^\rho{}_{\mu\nu}
  &= \frac{1}{2}h^{\rho\sigma}\left(\tau_\mu K_{\sigma\nu}+\tau_\nu K_{\sigma\mu}+L_{\sigma\mu\nu}\right)\,.
\end{align}
where $K_{\mu\nu}$ and $L_{\sigma\mu\nu}$ satisfy additional constraints (see~\cite{Hartong:2015zia}).

First, it is important to note that we no longer require the connection to be torsionless.
As a result, $d\tau$ is no longer necessarily zero, as we can see for example from the torsion of the connection $\check\Gamma^\rho_{\mu\nu}$ in~\eqref{eq:gamma-check-tnc},
\begin{equation}\label{eq:torsion2}
  2\check\Gamma^\rho_{[\mu\nu]}
  =-v^\rho\left(\partial_\mu\tau_\nu-\partial_\nu\tau_\mu\right)\,,
\end{equation}
Following \cite{Christensen:2013lma,Christensen:2013rfa,Figueroa-OFarrill:2020gpr} we distinguish three cases:
\begin{enumerate}
  \item Zero torsion $\pd_{[\mu}\tau_{\nu]}=0$ corresponds to `regular' Newton--Cartan (NC) geometry, which is the case we considered in Section~\ref{sec:nc-basics} above.
\item `Twistless' torsion, defined by
  \begin{equation}
  \label{eq:TTNC}
    \tau_{[\mu}\partial_\nu\tau_{\rho]}=0
    \quad\Longleftrightarrow\quad
    h^{\mu\rho}h^{\nu\sigma}\left(\partial_\rho\tau_\sigma-\partial_\sigma\tau_\rho\right)=0\,,
  \end{equation}
  which implies that $\tau_\mu$ can be used to define spatial hypersurfaces of co-dimension one, where $h_{\mu\nu}$ pulls back to a nondegenerate Riemannian metric.
  The corresponding geometry is known as twistless torsional Newton--Cartan (TTNC) geometry,
  and this will be our main focus in the following.
\item No constraint on $\pd_{[\mu}\tau_{\nu]}$ and general torsional Newton--Cartan (TNC) geometry.
\end{enumerate} 
The notion of TTNC geometry goes back to \cite{Julia:1994bs}, however in that work a conformal rescaling was done to go to a frame in which there is no torsion.
The benefit of adding torsion to the formalism, including the general TNC case, was first considered in \cite{Christensen:2013lma,Christensen:2013rfa}. One can also obtain TTNC geometries by gauging the Schr\"odinger algebra \cite{Bergshoeff:2014uea,Afshar:2015aku}.

In the case of TTNC geometry we have the following useful identities 
\begin{align}
  h^{\mu\rho}h^{\nu\sigma}\left(\partial_\rho a_\sigma-\partial_\sigma a_\rho\right)
  &= h^{\mu\rho}h^{\nu\sigma}\left(\nabla_\rho a_\sigma-\nabla_\sigma a_\rho\right)=0\,,
  \label{eq:TTNCidentity1}\\
  \partial_\mu \tau_\nu-\partial_\nu \tau_\mu
  &= a_\mu\tau_\nu-a_\nu\tau_\mu \,.
  \label{eq:TTNCidentity2}
\end{align}
in terms of the `acceleration' vector $a_\mu= \LL_v\tau_\mu$ of the foliation.
The second identity tells us that $h^{\mu\nu}a_\mu$ describes the TTNC torsion,
which is why the latter is sometimes also known as the torsion vector.
Equation~\eqref{eq:TTNCidentity1} tells us that the twist tensor vanishes, which is why we refer to the geometry as twistless torsional NC geometry.

Additionally, recall that the connection $\check\Gamma^\rho_{\mu\nu}$ is not invariant under Galilean boosts.
Using the Bargmann $U(1)$ field $m_\mu$, we can choose $K_{\mu\nu}$ and $L_{\mu\nu\rho}$ such that
the affine connection is invariant under Galilean boosts $\delta_G\Gamma^{\rho}_{\mu\nu}=0$, 
\begin{align}
  K_{\mu\nu}
  &= \partial_\mu m_\nu-\partial_\nu m_\mu\,,
  \label{eq:choiceK} 
  \\
  L_{\sigma\mu\nu}
  &= m_\sigma\left(\partial_\mu\tau_\nu-\partial_\nu\tau_\mu\right)
  - m_\mu\left(\partial_\nu\tau_\sigma-\partial_\sigma\tau_\nu\right)
  - m_\nu\left(\partial_\mu\tau_\sigma-\partial_\sigma\tau_\mu\right)\,.
  \label{eq:choiceL}
\end{align}
The resulting connection then takes the form
\begin{equation}\label{eq:barGammaTNC}
  \bar\Gamma^{\rho}_{\mu\nu}
  = -\hat v^\rho\partial_\mu\tau_\nu
  +\frac{1}{2}h^{\rho\sigma}\left(
    \partial_\mu\bar h_{\nu\sigma}+\partial_\nu \bar h_{\mu\sigma}-\partial_\sigma\bar h_{\mu\nu}
  \right)\,.
\end{equation}
which is the generalization of the connection~\eqref{eq:gammabar} to nonzero torsion.
This is not the unique boost-invariant connection, but it is one of the more natural choices and it is commonly used in the literature.
However, note that the connection~\eqref{eq:barGammaTNC} is no longer invariant under $U(1)$ transformations in the presence of torsion.%
\footnote{%
  As before, the connection~\eqref{eq:barGammaTNC} is not invariant under all local symmetries, so it is strictly speaking not an affine connection in the regular sense.
  In particular, not all of its curvature tensors will be invariant under local symmetries.
  However, this potential problem is circumvented by explicitly checking that all actions we consider are invariant under non-manifest gauge symmetries.
  Often, this is guaranteed by their origin as a limit or reduction of a Lorentz-invariant action.
}
With torsion, one can show that it is no longer possible to build a connection that is invariant under both $U(1)$ transformations and Galilean boosts.

Finally, following Appendix~\ref{sec:conventions}, we can define the Riemann curvature tensor associated with for example
the TNC connection~\eqref{eq:gamma-check-tnc} that is not invariant under boosts (but is invariant under $U(1)$ gauge transformations) 
or
the TNC connection~\eqref{eq:barGammaTNC}
which is boost-invariant (but not invariant under $U(1)$ gauge transformations)
through
\begin{align}
  \label{eq:check-riemann}
  \check{R}_{\mu\nu\sigma}{}^\rho
  &=-\partial_\mu\check\Gamma^\rho_{\nu\sigma}+\partial_\nu\check\Gamma^\rho_{\mu\sigma}
  -\check\Gamma^\rho_{\mu\lambda}\check\Gamma^\lambda_{\nu\sigma}
  +\check\Gamma^\rho_{\nu\lambda}\check\Gamma^\lambda_{\mu\sigma}\,,
  \\
  \label{eq:bar-riemann}
  \bar{R}_{\mu\nu\sigma}{}^\rho
  &=-\partial_\mu\bar\Gamma^\rho_{\nu\sigma}+\partial_\nu\bar\Gamma^\rho_{\mu\sigma}
  -\bar\Gamma^\rho_{\mu\lambda}\bar\Gamma^\lambda_{\nu\sigma}
  +\bar\Gamma^\rho_{\nu\lambda}\bar\Gamma^\lambda_{\mu\sigma}\,,
\end{align}
from which one can compute the Ricci tensor and curvature scalars in the usual way. 

\subsubsection{Type I TNC geometry from null reduction}
\label{sssec:type-I-tnc-geometry-null-reduction}
Another way to obtain $(d+1)$-dimensional type I Newton--Cartan geometry is through null reduction of a $(d+2)$-dimensional Lorentzian geometry (see also Section~\ref{sec:nullred}). 
Choosing coordinates $(u,x^\mu)$ such that the null isometry is generated by $\pd_u$, we can parametrize such a metric as
\begin{equation}
  \label{eq:nullreduction-metric}
  ds^2
  = g_{MN} dx^M dx^N
  = 2\tau_\mu (du - m_\mu dx^\mu) + h_{\mu \nu} dx^\mu dx^\nu  \,,
\end{equation}
where none of the metric components depend on $u$. The $x^\mu$ are $(d+1)$-dimensional coordinates.
The splitting of the $g_{\mu\nu}$ components into $(\tau_\mu,h_{\mu\nu},m_\mu)$ is ambiguous, which is the origin of the local Galilean boost transformation of $h_{\mu\nu}$ and $m_\mu$ in~\eqref{eq:type-I-tnc-metric-boost-tr}.
Alternatively, the metric and its inverse can be naturally decomposed in terms of the boost-invariant quantities defined in~\eqref{eq:type1-tnc-spatial-metric-and-inverse}, \eqref{eq:hatv}, \eqref{eq:barh} and~\eqref{eq:hatPhi},
\begin{equation}
  \label{eq:null-metric-and-inverse}
  g_{MN}
  = \begin{pmatrix}
    0 & \tau_\nu
    \\
    \tau_\mu & \bar{h}_{\mu\nu}
  \end{pmatrix},
  \qquad
  g^{MN}
  = \begin{pmatrix}
    2 \hat\Phi
    & - \hat{v}^\nu
    \\
    - \hat{v}^\nu
    & h^{\mu\nu}
  \end{pmatrix}.
\end{equation}
Furthermore, since there is no restriction on $\tau_{\mu}$, the resulting geometry will generically be torsionful.
The fact that null reduction of a Lorentzian geometry leads to type I Newton--Cartan geometry with local Bargmann symmetries can be understood from the fact that the Bargmann algebra can be obtained from a null reduction of the Poincaré algebra, as we mentioned around equation~\eqref{eq:null-centralizer-generators}.

\subsection{Particle action with type I TNC background}
\label{ssec:type-I-tnc-particle-action}
We have obtained type I TNC geometry from a gauging of the Bargmann algebra and from null reduction. We will next consider a third way of obtaining it, namely from a limit procedure. In the process, we will construct the TNC analogue of the Lorentzian point particle action\begin{equation}
  \label{eq:lor-pp-action}
  S = - mc \int d\lambda \sqrt{ - g_{\mu\nu} \dot{x}^\mu \dot{x}^\nu}
  + q \int A_\mu \dot{x}^\mu\,.
\end{equation}
First, we will consider the analogue of the İnönü--Wigner-type contraction~\eqref{eq:bargmann-contraction-defs} using a background Maxwell potential, following for example~\cite{Jensen:2014wha,Bergshoeff:2015uaa,Bergshoeff:2015sic}.
Next, we will obtain the same TNC action from a null reduction of a massless particle coupled to a Lorentzian background with a null isometry.

\subsubsection{From a contraction with Maxwell background}
\label{sssec:type-I-tnc-particle-action-limit}
In terms of the Poincaré algebra trivially extended with a $U(1)$ generator $Q$, the connection~\eqref{eq:gauging-poincare-connection-decomposition} becomes
\begin{align}\label{eq:redefcalA}
  \mathcal{A}_\mu
  &= E_\mu{}^A P_A + \frac{1}{2} \Omega_\mu{}^{AB} M_{AB}
  + A_\mu Q\nonumber
  \\
  &= \tau_\mu H
  + e_\mu^a P_a
  + \Omega_\mu{}^a G_a
  + \frac{1}{2} \Omega_\mu{}^{ab} J_{ab}
  + m_\mu N\,,
\end{align}
where we used the redefinition~\eqref{eq:bargmann-contraction-defs} to obtain the second line, and where we have defined
\begin{equation}
  \label{eq:bargmann-vielbein-u1-redef}
  E_\mu{}^0
  = c \tau_\mu + \frac{1}{c}m_\mu\,,
  \qquad
  E_\mu{}^a
  = e_\mu{}^a\,,
  \qquad
  A_\mu = \tau_\mu\,.
\end{equation}
The second line of \eqref{eq:redefcalA} corresponds to the Bargmann connection~\eqref{eq:gauging-bargmann-connection-decomposition} upon taking the $c\to\infty$ contraction.

Recalling that $g_{\mu\nu}=-E_\mu{}^0 E_\nu{}^0+\delta_{ab}E_\mu{}^a E_\nu{}^b$
and subsequently applying the parametrization above to the Lorentzian action~\eqref{eq:lor-pp-action},
we get
\begin{equation}
  \label{eq:type-I-tnc-pp-action-from-contraction}
  S = (q - mc^2) \int d\lambda \tau_\mu \dot{x}^\mu
  + \frac{m}{2} \int d\lambda 
  \frac{\bar{h}_{\mu\nu}\dot{x}^\mu\dot{x}^\nu}{\tau_\rho\dot{x}^\rho}
  + \OO(1/c^2),
\end{equation}
where $\bar{h}_{\mu\nu}$ is defined in~\eqref{eq:barh} and we recall that $h_{\mu\nu}=\delta_{ab}e_\mu{}^a e_\nu{}^b$.
We see that we can cancel the leading-order term by setting $q=mc^2$, corresponding to an extremal charge.
In the limit $c\rightarrow \infty$,
the remaining action
then describes the coupling of a point particle of mass $m$ to type I TNC geometry.
By construction, it is invariant under the Bargmann transformations~\eqref{eq:deltabar-bargman-trafos}, as can also be checked explicitly.
Additionally, note that it agrees with the action given in~\eqref{eq:geodesic}.

\subsubsection{From null reduction}
\label{sssec:type-I-tnc-particle-action-nr}
Finally, we can obtain the same action from a null reduction of the Lorentzian action for a massless particle without the Maxwell coupling.
Using the parametrization in~\eqref{eq:nullreduction-metric} for a $(d+2)$-dimensional Lorentzian metric $g_{MN}$ with a null isometry, the action for a massless particle is given by
\begin{equation}
  \label{eq:lor-pp-action-null-background}
  S
  = \int \frac{1}{2e}g_{MN}\dot X^M\dot X^N d \lambda
  = \int \left(
    \frac{1}{e}\dot{u}\tau_\mu \dot X^\mu
    + \frac{1}{2e}\bar h_{\mu\nu}\dot X^\mu \dot X^\nu
  \right) d\lambda\,.
\end{equation}
The momentum associated to the null direction $p_u = \pd \LL/ \pd \dot{u} = \tau_\mu \dot{X}^\mu / e$ is conserved due to the isometry,
which allows us to solve for the worldline `einbein' $e = \tau_{\mu}\dot{X}^\mu/ p_u$.
After setting $p_u = m$, the action~\eqref{eq:lor-pp-action-null-background} then reproduces
\begin{equation}
  \label{eq:type-I-tnc-pp-action-from-null-red}
  S
  = \frac{m}{2} \int \frac{\bar{h}_{\mu\nu} \dot{X}^\mu \dot{X}^\nu}{\tau_\rho \dot{X}^\rho} d\lambda\,,
\end{equation}
which is the same action we obtained from~\eqref{eq:type-I-tnc-pp-action-from-contraction} after canceling the leading-order term.

\subsection{Gravity action for type I TNC geometry}
\label{ssec:type-I-tnc-gravity-action}
Similarly to the point particle action, we can construct a gravitational action for dynamical type I torsional Newton--Cartan (TNC) geometry in two ways.
First, we will consider an İnönü--Wigner-type contraction of Einstein gravity coupled to a Maxwell action.
Next, we will perform the null reduction of the Einstein--Hilbert action on a Lorentzian background with a null isometry, which also gives an action that is invariant under Bargmann symmetries.
We then show that the two actions obtained in this way are in fact equal.
As far as we know, this relation has not yet been explicitly identified in the literature in this form, although a similar discussion appears in frame language in~\cite{simsek-thesis}.

\subsubsection{From a contraction of Einstein--Maxwell}
\label{sssec:type-I-tnc-gravity-action-limit}
In our first approach to constructing an action for dynamical type I TNC geometry, we start from Einstein--Maxwell gravity, whose Lagrangian is\footnote{The powers of $c$ follow from the fact that in our $c$-expansion we have $\sqrt{-g}=ce\left(1+\mathcal{O}(c^{-2})\right)$.}
\begin{equation}
  \LL_\text{EM}
  = \frac{c^3}{16\pi G} \sqrt{-g} R
  - \frac{1}{4 c\, k^2} \sqrt{-g}
  g^{\mu\rho} g^{\nu\sigma} F_{\mu\nu} F_{\rho\sigma}\,.
\end{equation}
As we will explain in more detail in Section~\ref{ssec:exp-action}, the Lorentzian metric $g_{\mu\nu}$ and the Levi-Civita Ricci scalar $R$ can be covariantly expanded in powers of $c^2$.
In particular, we will see that the leading-order term in the expansion of $R$ is not related to the curvature of a Newton--Cartan connection, but instead it depends on 
\begin{equation}
\tau_{\mu\nu}=2\pd_{[\mu}\tau_{\nu]}\,,
\end{equation}
which parametrizes the torsion of such a connection.
Specifically, we get
\begin{equation}
  \LL_\text{EM}
  = e h^{\mu\rho} h^{\nu\sigma}
  \left(
    \frac{c^6}{64\pi G} 
    \tau_{\mu\nu} \tau_{\rho\sigma}
    - \frac{1}{4 k^2}
    F_{\mu\nu}F_{\rho\sigma}
  \right)
  + \OO(c^4)\,.
\end{equation}
where $e=\det(\tau_\mu,e_\mu{}^a)$ is the Newton--Cartan vielbein determinant.
In analogy with the expansion of the point particle action in~\eqref{eq:type-I-tnc-pp-action-from-contraction}, we see that we can cancel this leading-order term by setting
\begin{equation}
  \frac{1}{k^2} = \frac{c^6}{16\pi G}\,,
  \qquad
  A_\mu = \tau_\mu\,.
\end{equation}
Once the leading-order terms are cancelled, the $c\to\infty$ limit gives
\begin{equation}
  \label{eq:type-one-action-limit}
  \LL
  = \frac{e}{16\pi G} \left(
    h^{\mu\nu} \check{R}_{\mu\nu}
    + \frac{1}{2} h^{\mu\nu} a_\mu a_\nu
    + \frac{1}{2} h^{\mu\rho} h^{\nu\sigma} \tau_{\mu\nu} m_{\rho\sigma}
  \right)\,,
\end{equation}
where we have defined 
\begin{equation}
m_{\mu\nu} = 2\pd_{[\mu}m_{\nu]}\,,
\end{equation}
and where $\check{R}_{\mu\nu}$ is the Ricci tensor associated to the connection~$\check{\Gamma}^\rho_{\mu\nu}$ in~\eqref{eq:gamma-check-tnc},
and we have rescaled $G\to G c^4$.

\subsubsection{From null reduction of Einstein--Hilbert}
\label{sssec:type-I-tnc-gravity-action-nr}
On the other hand, we can also obtain an action for type I Newton--Cartan gravity using a null reduction of the Einstein--Hilbert action.
Rewriting the $(d+2)$-dimensional Levi--Civita connection, which we denote by $\hat{\Gamma}^R_{MN}$, using the decomposition~\eqref{eq:null-metric-and-inverse} of the metric, we find
$\hat\Gamma^u_{uu} = \hat\Gamma^\rho_{uu} = 0$
and
\begin{subequations}
  \label{eq:lc-null-decomposition}
  \begin{align}
    \hat\Gamma^u_{\mu u}
    &= \frac{1}{2} \hat{a}_\mu\,,
    \\
    \hat{\Gamma}^u_{\mu\nu}
    &= - 2 \tau_{(\mu} \pd_{\nu)} \hat{\Phi}
    - \bar{K}_{\mu\nu}\,,
    \\
    \hat{\Gamma}^\rho_{\mu u}
    &= \frac{1}{2} \tau_{\mu\lambda} h^{\lambda\rho}\,,
    \\
    \hat{\Gamma}^\rho_{\mu\nu}
    &= - \hat{v}^\rho \pd_{(\mu} \tau_{\nu)}
    + \frac{1}{2} h^{\rho\lambda} \left[
      \pd_\mu \bar{h}_{\nu\lambda}
      + \pd_\nu \bar{h}_{\lambda\nu}
      - \pd_\lambda \bar{h}_{\mu\nu}
    \right]
    = \bar\Gamma^\rho_{(\mu\nu)},\,.
  \end{align}
\end{subequations}
Here, the boost-invariant objects $\hat a_\mu$ and $\bar K_{\mu\nu}$ are
\begin{equation}
  \hat{a}_\mu
  = \LL_{\hat{v}} \tau_\mu
  = \hat{v}^\rho \tau_{\rho\mu}\,,
  \qquad
  \bar{K}_{\mu\nu}
  = - \frac{1}{2} \LL_{\hat{v}} \bar{h}_{\mu\nu}\,,
\end{equation}
and $\bar\Gamma^\rho_{\mu\nu}$ is the boost-invariant connection defined in~\eqref{eq:barGammaTNC}.
Similarly, we can decompose the Ricci tensor $\hat{R}_{MN}$ into lower-dimensional components.
For example, we get
\begin{equation}
  \label{eq:Rhat-uu-cpt}
  \hat{R}_{uu}
  = \frac{1}{4} h^{\mu\rho} h^{\nu\sigma} \tau_{\mu\nu} \tau_{\rho\sigma}\,,
\end{equation}
which we will use later on.
Using the decomposition of the $(d+2)$-dimensional Ricci scalar, the Einstein--Hilbert action becomes
\begin{equation}
  \label{eq:null-reduced-eh-action}
  \LL
  = \frac{\hat{E}}{16\pi \hat{G}}
  g^{MN} R_{MN}
  = \frac{e}{16\pi G} \left(
    h^{\mu\nu} \bar{R}_{\mu\nu}
    + \frac{1}{2} h^{\mu\nu} \hat{a}_\mu \hat{a}_\nu
    - \frac{1}{2} \hat{\Phi}
    h^{\mu\rho} h^{\nu\sigma} \tau_{\mu\rho} \tau_{\nu\sigma}
  \right)\,.
\end{equation}
Since all of its components are independent of $u$, we can interpret this as an action for $(d+1)$-dimensional type I TNC geometry.
Indeed, one can check that it is also invariant under Bargmann transformations.

This action is not obviously the same as the action~\eqref{eq:type-one-action-limit} we obtained from a limit.
However, as in the case of the point particle, the two actions are in fact equal.
To see this, one can use the relation~\eqref{eq:curvature-shift-in-connection} to rewrite
\begin{subequations}
  \begin{align}
    \label{eq:ricci-h-contracted-check-to-bar}
    h^{\mu\nu}\check{R}_{\mu\nu}
    &= h^{\mu\nu}\bar{R}_{\mu\nu}
    - \frac{1}{2} \hat{\Phi} h^{\mu\rho}
    h^{\mu\rho} h^{\nu\sigma} \tau_{\mu\nu} \tau_{\rho\sigma}
    - \frac{1}{2} m_\mu h^{\mu\nu} \tau_{\nu\rho} h^{\rho\sigma}
    \tau_{\sigma\alpha} h^{\alpha\beta} m_\beta
    \\
    &{}\qquad\nonumber
    - \frac{1}{2} h^{\mu\rho} h^{\nu\sigma} \tau_{\mu\nu} m_{\rho\sigma}
    + h^{\mu\nu} a^\rho \tau_{\rho\mu} m_\nu\,,
    \\
    \label{eq:acceleration-square-difference}
    \frac{1}{2} h^{\mu\nu} a_\mu a_\nu
    &= \frac{1}{2} h^{\mu\nu} \hat{a}_\mu \hat{a}_\nu
    - h^{\mu\nu} a^\rho \tau_{\rho\mu} m_\nu
    + \frac{1}{2} m_\mu h^{\mu\nu} \tau_{\nu\rho} h^{\rho\sigma}
    \tau_{\sigma\alpha} h^{\alpha\beta} m_\beta\,,
  \end{align}
\end{subequations}
which means that
\begin{equation}
  h^{\mu\nu} \check{R}_{\mu\nu}
  + \frac{1}{2} h^{\mu\nu} a_\mu a_\nu
  + \frac{1}{2} h^{\mu\rho} h^{\nu\sigma} \tau_{\mu\nu} m_{\rho\sigma}
  =
  h^{\mu\nu} \bar{R}_{\mu\nu}
  + \frac{1}{2} h^{\mu\nu} \hat{a}_\mu \hat{a}_\nu
  - \frac{1}{2} \hat{\Phi}
  h^{\mu\rho} h^{\nu\sigma} \tau_{\mu\nu} \tau_{\rho\sigma}\,.
\end{equation}
This identity equates the action~\eqref{eq:null-reduced-eh-action} obtained from null reduction of Einstein--Hilbert to the action~\eqref{eq:type-one-action-limit} that we obtained from a contraction of the Einstein--Maxwell action.

Finally, note that we could also have performed the null reduction at the level of the $(d+2)$-dimensional Einstein equations,
\begin{equation}
  \label{eq:higher-dim-ee}
  \hat{G}_{MN} = 8\pi \hat{G}\, \hat{T}_{MN}\,,
\end{equation}
where $\hat{G}_{MN}$ and $\hat{T}_{MN}$ are the higher-dimensional Einstein tensor and Lorentzian energy-momentum tensor, respectively.
In this case, we would obtain one additional equation of motion
that does not follow from the variations of the null-reduced action~\eqref{eq:null-reduced-eh-action},
\begin{equation}
  \label{eq:null-red-missing-eom}
  \hat{G}^{uu}
  =  8\pi \hat{G}\, T^{uu}\,.
\end{equation}
This is due to the fact that $g_{uu}=0$ was fixed off-shell to obtain the action~\eqref{eq:type-I-tnc-pp-action-from-null-red}.
However, the fact that this equation is missing does not make the reduction inconsistent, since it turns out that the remaining equations of motion agree with the null reduction of the higher-dimensional Einstein equation. They form a closed set under Bargmann transformations (see e.g. \cite{Bergshoeff:2021tfn}).
Therefore, we can add the missing equation of motion~\eqref{eq:null-red-missing-eom} by hand to the equations obtained by null reduction of the action.

\subsection{No mass coupling to torsionless type I Newton--Cartan gravity}
\label{ssec:type-I-tnc-mass-coupling}
We now demonstrate that the type I Newton--Cartan gravity actions that we constructed above cannot be coupled to mass sources without turning on torsion.
This leads us to the conclusion that type I Bargmann symmetry is not appropriate for reproducing Newtonian gravity, which contains mass coupling but requires vanishing torsion to ensure that time is absolute.

First, let us consider the formulation of the action in~\eqref{eq:type-one-action-limit}, which contains the Newton--Cartan variables $(\tau_\mu, h_{\mu\nu})$ and the $U(1)$ Bargmann potential $m_\mu$ as dynamical fields.
We define the following energy-momentum and mass currents from the coupling of a matter Lagrangian,
\begin{equation}
  \delta\LL_\text{mat}
  = e \left(
    T^\mu_\tau \delta \tau_\mu
    + \frac{1}{2} T^{\mu\nu}_h \delta h_{\mu\nu}
    + T^\mu_m \delta m_\mu
  \right)\,,
\end{equation}
where $T^\mu_\tau$ is the energy current, $T^{\mu\nu}_h$ the mass-momentum tensor\footnote{This is defined up to terms proportional to $v^\mu v^\nu$ since $v^\mu v^\nu\delta h_{\mu\nu}=0$ as a result of $v^\mu h_{\mu\nu}=0$.}, and $T^\mu_m$ the mass current.
We will show that having a non-zero mass density
$\rho = - \tau_\mu T^\mu_m$
is incompatible with the Newtonian requirement of vanishing torsion.
Consider the variation of the type~I gravity action in~\eqref{eq:type-one-action-limit} with respect to $m_\mu$,
\begin{equation}
  \delta_m \LL
  = - \frac{e}{8\pi G} G^\mu_m \delta m_\mu
  \qiq
  G^\mu_m
  = \frac{1}{2e} \pd_\nu \left(
    e h^{\nu\rho} h^{\mu\sigma} \tau_{\rho\sigma}
  \right)\,.
\end{equation}
We can rewrite the contraction of this vacuum equation of motion with $\tau_\mu$ as follows,
\begin{align}
  e \tau_\mu G^\mu_m
  = - \frac{e}{2} \left(\pd_\nu \tau_\mu\right)
  h^{\nu\rho} h^{\mu\sigma} \tau_{\rho\sigma}
  = - \frac{e}{4} h^{\mu\rho} h^{\nu\sigma} \tau_{\mu\nu} \tau_{\rho\sigma}\,.
\end{align}
As a result, we see that the $\tau_\mu$ projection of the $m_\mu$ equation of motion for type I Newton--Cartan gravity gives
\begin{equation}
  \label{eq:mass-density-torsion-coupling}
  \tau_\mu G^\mu_m
  = - \frac{1}{4} h^{\mu\rho} h^{\nu\sigma} \tau_{\mu\nu} \tau_{\rho\sigma}
  =8\pi G\tau_\mu T^\mu_m= - 8\pi G\, \rho\,.
\end{equation}
This implies that the Newtonian zero torsion requirement $d\tau=0$ is incompatible with nonzero mass density $\rho\neq0$.
In fact, the situation is worse still: having nonzero mass density breaks the twistless torsion condition, and therefore no longer even allows us to define spatial hypersurfaces.

The same result can be obtained from the null reduction at the level of the equations of motion~\eqref{eq:higher-dim-ee}.
From the reduction of the energy-momentum tensor $\hat{T}_{MN}$, we can identify the type I TNC mass current as
\begin{equation}
  T^\mu_m
  =- \hat{T}^\mu{}_u\,,
\end{equation}
following for example Appendix A2 of~\cite{Hartong:2016nyx}.
After null reduction, we therefore get
\begin{equation}
  \tau_\mu \hat{G}^\mu{}_u
  = \hat{G}_{uu}
  = \hat{R}_{uu}
  = \frac{1}{4} h^{\mu\rho} h^{\nu\sigma} \tau_{\mu\nu} \tau_{\rho\sigma}
  = \tau_\mu \hat{T}^\mu{}_u
  = -\rho\,,
\end{equation}
where we have used Equation~\eqref{eq:Rhat-uu-cpt}.
This reproduces~\eqref{eq:mass-density-torsion-coupling} above.

Either way, we conclude that a different notion of dynamical Newton--Cartan geometry appears to be necessary to reproduce Newtonian gravity.
We address this problem in the next section.

\section{Non-relativistic expansion of general relativity}\label{sec:NRexpGR}
We will now derive an action whose equations of motion contain the Poisson equation of Newtonian gravity.
This construction requires a new notion of torsional Newton–Cartan (TNC) geometry based on an underlying symmetry algebra that differs from the usual Bargmann algebra.
This geometry naturally arises in a covariant $1/c$ expansion of general relativity, with $c$ being the speed of light.
The truncation of this expansion at subleading order provides the fields and transformation rules of `type II' TNC geometry.
The corresponding action and equations of motion
include the Poisson equation when sourced with non-relativistic matter.
More generally, they go beyond Newtonian gravity as
they allow for the effect of gravitational time dilation due to strong gravitational fields. 
The following is mainly based on Refs.~\cite{Hansen:2019svu,Hansen:2020pqs}, which built on earlier work by Dautcourt~\cite{Dautcourt:1996pm}
and crucially also on the more recent work by Van den Bleeken~\cite{VandenBleeken:2017rij}.

\subsection{Pre-non-relativistic form of general relativity}

In Lorentzian geometry, the slope of the light-cone is $1/c$, with $c$ denoting the speed of light.
The distinguishing feature of non-relativistic geometry is that this light-cone is flattened out completely.
This means that, in order to relate Lorentzian geometry to non-relativistic geometry, we need to perform an expansion around $c=\infty$.

The constant $c$ is of course dimensionful, but if we assume analyticity in $1/c$ of an appropriate set of variables then there must exist some other characteristic velocity that is small compared to $c$. However, the nature of this velocity is context dependent and can only be identified on shell for a specific problem. Nevertheless it is possible to formulate the general theory of the $1/c$ expansion without knowing what the dimensionless ratio(s) is/are in which we expand the Einstein equation. In this section, we focus on expanding in even powers of $1/c$, but we briefly discuss some features of the full expansion including odd powers in Section~\ref{ssec:odd}.

A convenient starting point for this expansion is the pre-non-relativistic (PNR) rewriting of GR,
where we make the way in which the speed of light enters in a Lorentzian metric manifest,
\begin{equation}
  \label{eq:tangent-metric-pnr-decomposition}
  g_{\mu\nu}
  = - c^2 T_\mu T_\nu + \Pi_{\mu\nu}\,,
  \qquad
  g^{\mu\nu}
  = - \frac{1}{c^2} V^\mu V^\nu + \Pi^{\mu\nu}\,.
\end{equation}
In this parametrization, the Lorentzian metric and its inverse are split into a component involving a `timelike' one-form or vector $T_\mu$ and~$V^\mu$ respectively, as well as the `spatial' symmetric tensors $\Pi_{\mu\nu}$ or $\Pi^{\mu\nu}$.
The latter two can be
written in terms of space-like vielbeine as  $\Pi^{\mu\nu}=\delta^{ab}E^\mu{}_a E^\nu{}_b$ and
$ \Pi_{\mu\nu}=\delta_{ab}E_\mu{}^a E_\nu^b$, where $a,b = 1 \ldots d$ are spacelike indices 
in the tangent space, and the total spacetime dimension is $d+1$. 

These PNR variables satisfy the following orthonormality relations,
\begin{equation}
  \label{eq:tangent-metric-pnr-orthogonality-completeness}
  T_\mu V^\mu = -1\,,
  \quad
  T_\mu \Pi^{\mu\nu} = 0\,,
  \quad
  \Pi_{\mu\nu} V^\nu = 0\,,
  \quad
  \delta^\mu_\nu = - V^\mu T_\nu + \Pi^{\mu\rho} \Pi_{\rho\nu}\,.
\end{equation}
One can view the PNR  parametrization as a split of the tangent bundle in `temporal' and `spatial' components,
which also makes the factors of $c^2$ that appear in the Lorentzian metric explicit.
The new PNR tensor variables can then be expanded uniformly in $\sigma=1/c^2$, with the leading components being order one.

The next step is to obtain the form of the Einstein-Hilbert (EH) Lagrangian in terms of the PNR variables.
After some algebra,
it can be shown that (up to total derivatives) this takes the form%
\footnote{%
  The overall factor of $c^6$ arises from a combination of a factor of $c^2$ from the rewriting of the Levi--Civita Ricci scalar to the PNR Ricci scalar and a factor of $c^4$ from the dimensional prefactors of the action and the square root of the metric determinant.
}
\begin{equation}\label{eq:explicitsigma}
  \mathcal{L}_\text{EH}
  = \frac{c^6}{16\pi G} E\left[
    \frac{1}{4}\Pi^{\mu\nu}\Pi^{\rho\sigma}T_{\mu\rho}T_{\nu\sigma}
    +\sigma\Pi^{\mu\nu}\os{R}{C}_{\mu\nu}
    -\sigma^2 V^\mu V^\nu\os{R}{C}_{\mu\nu}
  \right]\,.
\end{equation}
where $T_{\mu\nu} = 2\partial_{[\mu } T_{\nu]} $ and $E = {\rm det} (T_\mu, E_\mu{}^a) $. 
The Ricci tensor $ \os{R}{C}_{\mu\nu} $ is defined in the usual way from the Riemann tensor $\os{R}{C}_{\mu\nu\rho}{}^\sigma$, see App.~\ref{sec:conventions} for our conventions.
However, the latter is now constructed from a PNR covariant derivative $\os{\nabla}{C}_\mu$ corresponding to a PNR connection $C^\rho_{\mu\nu}$, instead of the usual Levi--Civita connection.
The PNR connection is given by
\begin{equation}\label{eq:C_connection_notexpanded}
    C^{\rho}_{\mu\nu}
    = -V^\rho\partial_\mu T_\nu
    + \frac{1}{2}\Pi^{\rho\sigma}\left(
      \partial_\mu\Pi_{\nu\sigma}+\partial_\nu\Pi_{\mu\sigma}-\partial_\sigma\Pi_{\mu\nu}
    \right)\,,
\end{equation}
This connection satisfies
$\os{\nabla}{C}_\mu T_\nu = 0$
and
$\os{\nabla}{C}_\mu\Pi^{\nu\rho}=0$,
which are the PNR
analogues of Newton--Cartan metric compatibility conditions~\eqref{eq:nc-metric-compatibility}.
In addition, it satisfies
\begin{equation}
  \os{\nabla}{C}_\mu T^\nu
  = \frac{1}{2}\Pi^{\nu\rho}\mathcal{L}_V\Pi_{\rho\mu}\,,
  \qquad
  \os{\nabla}{C}_\mu \Pi_{\nu\rho}
  = T_{(\nu}\mathcal{L}_V\Pi_{\rho)\mu}\,.
\end{equation}
where $\mathcal{L}_V$ is the Lie derivative with respect to $V^\mu$.
Note that this new connection is in general
torsionful, since
$C^{\rho}_{[\mu\nu]} = - V^\rho \partial_{[\mu} T_{\nu]}$
is not necessarily zero.

From a Lorentzian point of view, the parametrization~\eqref{eq:explicitsigma} of the Einstein--Hilbert Lagrangian may seem odd, even though it is still invariant under local Lorentz symmetry.
Instead, we have written it in terms of variables that are adapted to the local Galilei symmetry that arises in the large speed of light expansion.%
\footnote{Similarly, a variant of this rewriting of the Einstein--Hilbert action with a slightly different adapted connection~\cite{Bekaert:2015xua,Hartong:2015xda} can be used to perform a small speed of light expansion \cite{Hansen:2021fxi} leading to Carroll or ultra-local gravity.} 
For more details on this, we refer the reader to \cite{Hansen:2020wqw}, where a novel Palatini-type formulation of GR is obtained that provides a natural starting point for a first-order non-relativistic expansion.
This involves  a reformulation of the Lorentzian Palatini action in terms of moving frames that exhibit local Galilean covariance in a large speed of light expansion.
Comparing Lorentzian and Newton-Cartan metric-compatibility then gives another explanation of the generic appearance of torsion in the non-relativistic expansion. 

We now wish to consider the large speed of light expansion of the Einstein--Hilbert action.
For this, it is useful to first consider the expansion of a general Lagrangian.

\subsection{Large speed of light expansion of general Lagrangians \label{sec:Lagrangians}}
Consider a Lagrangian $\mathcal{L}=\mathcal{L}(\sigma,\phi^I,\partial_\mu\phi^I)$ that is a function of some set of field $\phi^I(x;\sigma)$ and its derivatives, where we also allow for an explicit dependence on the speed of light.
The starting assumption is that, up to an overall power of $c$ which will be factored out, any field $\phi^I(x;\sigma)$
is analytic in $\sigma$ such that 
it admits a Taylor expansion around $\sigma=0$,
\begin{equation}\label{eq:expansion_field_general}
  \phi^I(x;\sigma)
  = \phi^I_{(0)}(x) + \sigma\phi^I_{(2)}(x) + \sigma^2\phi^I_{(4)}(x)
  + \order{\sigma^3}\,,
\end{equation}
where $\phi^I_{(n)}(x)$ is used to denote the coefficient of $c^{-n}$ in the expansion
and where $I$ a shorthand for any spacetime and/or internal indices.
We remind the reader that we restrict here to the case of even powers in $c$ only. 
Given this expansion of the fields, we want to expand the Lagrangian in powers of $\sigma$.
Note that the $\sigma$ dependence can come from the expansion of the background metric or matter fields, but also from parameters appearing in the kinetic or potential terms.
The result \cite{Hansen:2018ofj,Hansen:2019svu,Hansen:2020pqs} is that
\begin{equation}\label{eq:expLagrangian}
  \mathcal{L}(c^2,\phi,\partial_\mu\phi)
  = c^N\;\oss{\mathcal{L}}{-N}{LO}
  +c^{N-2}\;\;\oss{\mathcal{L}}{2-N}{NLO}
  +c^{N-4}\;\;\oss{\mathcal{L}}{4-N}{NNLO}
  +\order{c^{N-6}}\,,
\end{equation}
where we have taken the overall power of the Lagrangian to be $\sigma^{-N/2}=c^N$ for some~$N$.

Restricting for simplicity to the structure of the expanded Lagrangian for a single field, one finds the following LO and NLO terms,
\begin{align}
  \oss{\mathcal{L}}{-N}{LO}
  &= \tilde{\mathcal{L}}(0)
  = \oss{\mathcal{L}}{-N}{LO}(\phi_{(0)},\partial_\mu\phi_{(0)})\,,\label{eq:expLagrangian1}
  \\
  \oss{\mathcal{L}}{2-N}{NLO}
  &= \tilde{\mathcal{L}}'(0)
  = \left.\frac{\partial\tilde{\mathcal{L}}}{\partial\sigma}\right|_{\sigma=0}
    +\phi_{(2)}\frac{\partial\;\oss{\mathcal{L}}{-N}{LO}}{\partial\phi_{(0)}}
    +\partial_\mu\phi_{(2)}\frac{\partial\;\oss{\mathcal{L}}{-N}{LO}}{\partial\partial_\mu\phi_{(0)}}\nonumber
    \\
    &=\left.\frac{\partial\tilde{\mathcal{L}}}{\partial\sigma}\right|_{\sigma=0}+\phi_{(2)}\frac{\delta\;\oss{\mathcal{L}}{-N}{LO}}{\delta\phi_{(0)}}\,.\label{eq:expLagrangian2}
\end{align}
We observe that varying the NLO action with respect to the NLO field gives rise to the equations of motion of the LO field in the LO action.
A similar property holds at any order. 
The corresponding expression for the NNLO action can be found in \cite{Hansen:2019svu,Hansen:2020pqs}.
Following the general principle discussed above, its EOM satisfy the relations
\begin{equation}\label{eq:NNLO_Nvar_NLO_var}
  \frac{\delta\;\;\oss{\mathcal{L}}{4-N}{NNLO}}{\delta\phi_{(2)}}
  =\frac{\delta\;\;\oss{\mathcal{L}}{2-N}{NLO}}{\delta\phi_{(0)}}\,,
  \qquad
  \frac{\delta\;\;\oss{\mathcal{L}}{4-N}{NNLO}}{\delta\phi_{(4)}}
  = \frac{\delta\;\;\oss{\mathcal{L}}{2-N}{NLO}}{\delta\phi_{(2)}}
  =\frac{\delta\;\;\oss{\mathcal{L}}{-N}{LO}}{\delta\phi_{(0)}}\,.
\end{equation}
This general expansion can be applied to the spacetime fields of Lorentzian gravity as well as other types of (bosonic) Lorentzian fields that couple to Lorentzian geometry. 

\subsection{Type II TNC geometry} 
\label{ssec:type-II-tnc-geom}
We now show how this general procedure can be used to expand the fields describing a $(d+1)$-dimensional Lorentzian manifold.
We focus on the metric description, but we will also briefly comment on the vielbein description. 
By assumption, the PNR fields $T_\mu$ and $\Pi_{\mu\nu}$ in the parametrization~\eqref{eq:tangent-metric-pnr-decomposition} of the Lorentzian metric are taken to be analytic in $\sigma$,
and thus they admit a Taylor expansion.
This means we can write them as
\begin{subequations}
  \label{eq:pnr-down-expansion}
  \begin{align}
    T_\mu
    &= \tau_\mu+c^{-2}m_\mu+c^{-4}B_\mu+\order{c^{-6}}\,,
    \label{eq:Tmu}
    \\
    \Pi_{\mu\nu}
    &= h_{\mu\nu}+c^{-2}\Phi_{\mu\nu}+c^{-4}\psi_{\mu\nu}+\order{c^{-4}}\,,
  \end{align}
\end{subequations}
and similarly for the inverse fields
\begin{subequations}
  \label{eq:pnr-up-expansion}
  \begin{align}
    \label{Tup}
    V^\mu
    &= v^\mu+c^{-2}\left(v^\mu v^\rho m_\rho
    -h^{\mu\rho}v^\sigma \Phi_{\rho\sigma}\right)
    +\order{c^{-4}}\,,
    \\ 
    \label{Piup}
    \Pi^{\mu\nu}
    &= h^{\mu\nu}
    +c^{-2}\left(2h^{\rho(\mu}v^{\nu)}m_\rho - h^{\mu\rho}h^{\nu\sigma}\Phi_{\rho\sigma} \right)
    +\order{c^{-4}}\,.
  \end{align}
\end{subequations}
The various tensors in this expansion satisfy orthonormality conditions that follow from expanding~\eqref{eq:tangent-metric-pnr-orthogonality-completeness}, which can be found in \cite{Hansen:2020pqs}.
In particular, we can use these relations to solve for the subleading fields in~\eqref{eq:pnr-down-expansion} in terms of the fields in~\eqref{eq:pnr-up-expansion}.

Up to next-to-leading order, the expansion therefore features the following fields
\begin{equation}
  \mbox{LO fields}:   \quad  \tau_\mu \,, h_{\mu \nu}\,,
  \qquad
  \mbox{NLO fields}:  \quad m_\mu \,, \Phi_{\mu \nu} \,,
\end{equation}
along with their inverses.
As we will show momentarily, the LO fields exhibit precisely the Galilean transformation rules
of the corresponding fields with the same names in type I TNC geometry.
In a slight abuse of notation, we have also used the same name $m_\mu$ as in type I for one of the NLO fields, but this field will in general transform {\it differently} in the current notion of torsional Newton--Cartan geometry.
Finally, we now also have an extra field $\Phi_{\mu \nu}$ at NLO.

Along with their transformations,
the collection of these four fields
is what defines \emph{type II TNC geometry}.
As we will see below, this NLO geometric structure in fact allows us to describe both the NLO and NNLO actions and equations of motion of non-relativistic gravity due to the simple form of the LO contributions.

We can find  the transformation rules of the expanded fields
by performing a similar large $c$ expansion
in the transformations of the Lorentzian metric and vielbeine under diffeomorphisms and local Lorentz transformations.
This leads to the following transformations of the LO and NLO fields,
\begin{subequations}
  \label{eq:symsTNCII}
  \begin{align}
    \bar\delta\tau_\mu
    &= \mathcal{L}_\xi\tau_\mu\,,
    \label{eq:trafo5a}
    \\
    \delta h_{\mu\nu}
    &= \mathcal{L}_\xi h_{\mu\nu}+\tau_\mu\lambda_\nu+\tau_\nu\lambda_\mu\,,
    \label{eq:trafo5}
    \\
    \bar\delta m_\mu
    &= \mathcal{L}_\xi m_\mu+\lambda_\mu+\partial_\mu \Lambda -\Lambda a_\mu+h^{\rho\sigma}\zeta_\sigma\left(\partial_\rho\tau_\mu-\partial_\mu\tau_\rho\right)\,,
    \label{eq:torsional_u(1)_1}
    \\
    \bar\delta \Phi_{\mu\nu}
    &= \mathcal{L}_\xi \Phi_{\mu\nu}+2\lambda_a\left(\tau_{(\mu}\pi^a_{\nu)}+m_{(\mu}e^a_{\nu)}\right)+2\eta_a e^a_{(\mu}\tau_{\nu)}+2\Lambda K_{\mu\nu}
    +2\check{\nabla}_{(\mu}\zeta_{\nu)}\,,
    \label{eq:torsional_u(1)_2}
  \end{align}
\end{subequations}
where $\lambda_\mu= e_\mu^a \lambda_a$ is the Galilean boost parameter which obeys $v^\mu\lambda_\mu=0$ and the parameter $\eta_a$ arises from subleading boost.
Additionally, the notation $\bar\delta$ foreshadows a link to the transformations~\eqref{eq:bardeltaA-def} in the gauging procedure, and the fields $e_\mu{}^a$ and $\pi_\mu{}^a$ are the leading and subleading terms in the expansion of the spatial vielbeine $E_\mu{}^a$, respectively.
We similarly expanded the diffeomorphisms so that $\xi^\mu$ is the LO part, while  the subleading diffeomorphisms are $\zeta^\mu$ which we have decomposed above as 
\begin{equation}
  \zeta^\mu = -\Lambda v^\mu+h^{\mu\nu}\zeta_\nu\,.
  \label{eq:torsionalU1_subleading_diffeo}
\end{equation}
In these expressions, we used the acceleration
and extrinsic curvature tensors
\begin{equation}
  a_\mu
  =\mathcal{L}_v \tau_\mu
  = v^\nu \tau_{\mu\nu}\,,
  \qquad
  K_{\mu\nu}
  = -\frac{1}{2}\mathcal{L}_v h_{\mu\nu}\,,
\end{equation}
which will also appear in the expanded EH action. 
We note that it can be easily checked that the transformations of the 
three fields $\tau_\mu$, $h_{\mu \nu}$ and $m_\mu$ in~\eqref{eq:symsTNCII} reduce to those of the corresponding type I TNC fields in \eqref{eq:type-I-tnc-metric-boost-tr} when $\tau_\mu$ is closed.

To write the expanded Einstein--Hilbert action, we need to introduce a connection on type II TNC geometry.
At leading order in the expansion of $C^\rho_{\mu\nu}$ in \eqref{eq:C_connection_notexpanded}, we recover the torsionful connection
$\check\Gamma^\rho_{\mu\nu}$ from~\eqref{eq:gamma-check-tnc},
\begin{equation}
  \label{eq:GammatypeII}
  \check\Gamma^\rho_{\mu\nu}
  = \left.C^\rho_{\mu\nu}\right|_{\sigma=0}
  = -v^\rho\partial_\mu\tau_\nu
  + \frac{1}{2}h^{\rho\sigma}\left(
    \partial_\mu h_{\nu\sigma}+\partial_\nu h_{\mu\sigma}-\partial_\sigma h_{\mu\nu}
  \right)\,.
\end{equation}
This combination is in some sense the minimal collection of terms that transforms as an affine connection under diffeomorphisms. Moreover, it follows from the metric compatibility conditions on the 
PNR connection $C^\rho_{\mu\nu}$  that 
$\check\Gamma^\rho_{\mu\nu}$ is a Newton--Cartan metric compatible connection satisfying the properties 
$\nabla_\mu \tau_\nu=0$ and $\nabla_\mu h^{\nu \rho}=0 $.
It does transform under local Galilei boosts, and therefore the corresponding curvature tensors will generically transform under boosts as well.
However, if we start from an action that is invariant under Lorentz boosts, such as the Einstein--Hilbert action, the expansion will only produce scalar combinations which are invariant under Galilean boosts.

\subsubsection*{Type II TNC symmetry algebra and Lie algebra expansion}
In Section~\ref{sssec:type-I-tnc-geometry-gauging-bargmann}, we derived the transformation properties~\eqref{eq:deltabar-bargman-trafos} of the type I TNC fields from the gauging of the Bargmann algebra.
It turns out that the transformations~\eqref{eq:symsTNCII} of the type II TNC fields can likewise be obtained from the gauging of an algebra.
In fact, the corresponding symmetry algebra follows from an expansion of the Poincar\'e algebra using the general method of Lie algebra expansions.%
\footnote{This method has been considered in for example~\cite{deAzcarraga:2002xi,Izaurieta:2006zz,Khasanov:2011jr} 
and was applied to the  $1/c^2$ expansion of the Poincar\'e algebra in 
\cite{Hansen:2019vqf,Hansen:2020pqs} as well as in \cite{Bergshoeff:2019ctr,Gomis:2019fdh,Gomis:2019sqv,deAzcarraga:2019mdn}.}

Applying the pre-non-relativistic decomposition to the Poincar\'e-valued Cartan connection~\eqref{eq:gauging-poincare-connection-decomposition} we get
\begin{equation}
  \mathcal{A}_\mu
  = H T_\mu+P_a E_\mu^a+B_a\Omega_\mu{}^a+\frac{1}{2}J_{ab}\Omega_\mu{}^{ab}\,,
\end{equation}
which contains the relativistic vielbeine $T_\mu$ and  $E_\mu^a$ along with the boost connection $\Omega_\mu{}^a$ and the rotation connection $\Omega_\mu{}^{ab}$.
If we schematically write this Cartan connection as $\mathcal{A}_\mu=T_I\mathcal{A}_\mu^I$ and expand its components as $\mathcal{A}_\mu^I=\sum_{n=0}^\infty \sigma^n\os{\mathcal{A}}{2n}_\mu^I$
we obtain the new generators $T^{(n)}_I= T_I \otimes \sigma^n$, where $n\ge 0$ will be referred to as the level. 

Using this expansion of the generators $T_I^{(n)}$, one obtains an algebra whose nonzero commutation relations are~\cite{Hansen:2019vqf}
\begin{align}\label{eq:commutation_relations_ring}
  \left[H^{(m)},B^{(n)}_a\right]
  &= P^{(m+n)}_a\,,
  \quad
  \left[P^{(m)}_a,B^{(n)}_b\right]
  = \delta_{ab}H^{(m+n+1)}\,,
  \quad
  \left[B^{(m)}_a,B^{(n)}_b\right]
  = -J^{(m+n+1)}_{ab}\,,
  \nonumber\\
  \left[J^{(m)}_{ab},P^{(n)}_c\right]
  &= \delta_{ac}P_b^{(m+n)}-\delta_{bc}P_a^{(m+n)}\,,
  \quad
  \left[J^{(m)}_{ab},B^{(n)}_c\right]
  = \delta_{ac}B_b^{(m+n)}-\delta_{bc}B_a^{(m+n)}\,,
  \nonumber\\
  \left[J_{ab}^{(m)},J_{cd}^{(n)}\right]
  &= \delta_{ac}J_{bd}^{(m+n)}-\delta_{bc}J_{ad}^{(m+n)}-\delta_{ad}J_{bc}^{(m+n)}+\delta_{bd}J_{ac}^{(m+n)}\,.
\end{align}
We can quotient out all generators with level $n>L$ for some $L$, which amounts to truncating the $1/c^2$ expansion.
At the lowest level level $L=0$, the algebra is isomorphic to the Galilei algebra and the gauging of the algebra can be shown to generate the transformation rules of $\tau_\mu$ and $h_{\mu \nu}$. 
At the next level $L=1$, the number of generators doubles and we get a novel algebra which can be shown to lead to the full set of symmetries in~\eqref{eq:symsTNCII}, which now acts on all of the LO and NLO fields.

Note that the $L=1$ truncation of the algebra~\eqref{eq:commutation_relations_ring} does not have the Bargmann algebra as a subalgebra, since the would-be Bargmann extension $H^{(1)}$ is not central as it has non-zero commutator with the Galilean boosts $B^{(1)}_a$. We can obtain the Bargmann algebra as a quotient of the $L=1$ algebra by the ideal spanned by $J^{(1)}_{ab}, B^{(1)}_a, P^{(1)}_a$.
Whereas the Bargmann algebra encodes the local symmetries of type I TNC geometry, this algebra can be considered the `type II Bargmann algebra' associated to type II TNC geometry. 
As we discussed below~\eqref{eq:symsTNCII}, the corresponding transformations agree on torsionless geometries, but they are generically distinct.

\subsection{Expanding the Einstein--Hilbert action} 
\label{ssec:exp-action}
We now have all ingredients in place to find the LO, NLO and NNLO terms in the expansion of the Einstein--Hilbert (EH) action following the methods outlined in Section~\ref{sec:Lagrangians}.
Following the pre-non-relativistic (PNR) parametrization~\eqref{eq:tangent-metric-pnr-decomposition} and the expansion~\eqref{eq:pnr-down-expansion}, we therefore obtain a theory that is expressed in terms of the LO and NLO fields 
\begin{equation}
  \phi^I_{(0)}=\{\tau_\mu,\,h_{\mu\nu}\}\,,
  \qquad
  \phi^I_{(2)}=\{m_\mu,\,\Phi_{\mu\nu}\}\,.
\end{equation}
The NNLO fields $\phi^I_{(4)}=\{B_\mu,\,\psi_{\mu\nu}\}$ also enter in the NNLO EH action, but as we will see below, they only play a limited role, implementing a particular constraint.

Following the general form~\eqref{eq:expLagrangian}, the $1/c^2$ expansion of the EH Lagrangian reads
\begin{equation}
  \mathcal{L}_{\text{EH}}
  = c^6\left(\oss{\mathcal{L}}{-6}{\text{LO}}
  + \sigma\;\oss{\mathcal{L}}{-4}{\text{NLO}}
  + \sigma^2\; \oss{\mathcal{L}}{-2}{\text{NNLO}}
  +\mathcal{O}(\sigma^3)\right)\,.
\end{equation}
To compute the first terms in this expansion, we use the PNR form \eqref{eq:explicitsigma} of the EH action, the large $c$ expansions of the metric variables in~\eqref{eq:pnr-down-expansion} as well as the choice of connection in \eqref{eq:GammatypeII}.
We review the main results below, referring to~\cite{Hansen:2020pqs} for more detail.

The LO action is
\begin{equation}
  \label{eq:LO-action}
  \oss{\mathcal{L}}{-6}{LO}
  = \frac{e}{64\pi G}h^{\mu\nu}h^{\rho\sigma}\tau_{\mu\rho}\tau_{\nu\sigma}\,,
\end{equation}
where $e ={\rm det}( \tau_\mu, e^a_\mu) $ and $\tau_{\mu\nu} = \partial_\mu\tau_\nu-\partial_\nu\tau_\mu$ as usual.
We note that this LO action is manifestly invariant under Galilean boosts. 
Its variation takes the form
\begin{equation}
  \delta\oss{\mathcal{L}}{-6}{LO}
  = -\frac{1}{8\pi G}e\left(
    \os{G}{-6}^\alpha_{\tau}\delta \tau_{\alpha}
    + \frac{1}{2}\os{G}{-6}^{\alpha\beta}_{h}\delta h_{\alpha\beta}
  \right)\,,
\end{equation}
where the leading order equations of motion are
\begin{subequations}
  \label{eq:LOeoms} 
  \begin{align}
    \os{G}{-6}^{\alpha\beta}_{h}
    &= - \frac{1}{8}h^{\mu\nu}h^{\rho\sigma}\tau_{\mu\rho}\tau_{\nu\sigma}h^{\alpha\beta}
    + \frac{1}{2}h^{\mu\alpha}h^{\nu\beta}h^{\rho\sigma}\tau_{\mu\rho}\tau_{\nu\sigma}\,,\\
    \os{G}{-6}^\alpha_{\tau}
    &= \frac{1}{8}h^{\mu\nu}h^{\rho\sigma}\tau_{\mu\rho}\tau_{\nu\sigma}v^\alpha
    + \frac{1}{2}a_\mu h^{\mu\nu} h^{\rho\alpha}\tau_{\nu\rho}
    + \frac{1}{2}e^{-1}\partial_\mu\left(eh^{\mu\nu}h^{\rho\alpha}\tau_{\nu\rho}\right)\,.
  \end{align}
\end{subequations}
It can be shown that these equations imply $h^{\mu\nu}h^{\rho\sigma}\tau_{\mu\rho}\tau_{\nu\sigma}=0$.
Since the latter is a sum of squares, it implies $h^{\mu\nu}h^{\rho\sigma}\tau_{\mu\rho}=0$ or equivalently $\tau\wedge d\tau=0$, which is the twistless torsion (TTNC) condition discussed in Section~\ref{sssec:type-I-tnc-geometry-connection-curvature}.
Hence, the on-shell geometry arising from the expansion is a TTNC geometry \cite{Christensen:2013lma}.
Note that the LO equations of motion~\eqref{eq:LOeoms} vanish identically once the TTNC condition is imposed.

The NLO action then takes the form
\begin{equation}\label{eq:NLOLag}
  \oss{\mathcal{L}}{-4}{NLO}
  = - \frac{e}{8\pi G}\left(
    - \frac{1}{2}h^{\mu\nu}\check R_{\mu\nu}
    + \os{G}{-6}_\tau^\mu\, m_\mu
    + \frac{1}{2}\os{G}{-6}_h^{\mu\nu}\Phi_{\mu\nu}
  \right)\,.
\end{equation}
where $\os{G}{-6}^\mu_\tau$ and $\os{G}{-6}_h^{\mu\nu}$ are the LO EOMs given in~\eqref{eq:LOeoms}.
Since the latter are equivalent to the TTNC condition, we can also write the NLO action as the first term in \eqref{eq:NLOLag} together with a Lagrange multiplier term $\chi_{\rho \sigma}h^{\rho\mu}h^{\sigma\nu} \tau_{\mu \nu}$ enforcing the TTNC condition.%
\footnote{Equivalently, one can obtain the resulting action by rewriting the first term in the PNR form of the EH action in terms of an auxiliary field  $\chi_{\mu \nu}$, with appropriate quadratic and linear terms, and then taking the large~$c$ limit, in the same way that the magnetic Carroll limit of GR was obtained in \cite{Hansen:2021fxi}.} 
The resulting NLO Lagrangian also has Galilean symmetries, and we will refer to its dynamics as Galilean gravity.
This theory was also studied in \cite{Bergshoeff:2017btm} using a first-order formalism.
Equation \eqref{eq:NLOLag} can be related to the Lagrangian appearing in that work by a specific choice of the undetermined Lagrange multipliers.

Following the general observation around Equation~\eqref{eq:expLagrangian2}, the leading order equations of motion are included in the NLO Lagrangian as the equations of motion of the NLO fields $m_\mu$ and $\Phi_{\mu\nu}$.
The NLO equations of motion of the LO fields $\tau_\mu$ and $h_{\mu\nu}$ are
\begin{subequations}
  \begin{align}
    \hspace{-.7cm}\os{G}{-4}_{\tau}^\nu
    &= \frac{1}{2}\left[
      2\left(
        h^{\mu\rho}h^{\nu\sigma}
        - h^{\mu\nu}h^{\rho\sigma}
      \right)
      \check{\nabla}_\mu K_{\rho\sigma}
      + v^\nu h^{\rho\sigma}\check R_{\rho\sigma}
      + \left(\check{\nabla}_\mu+2a_\mu \right)h^{\mu\rho}h^{\nu\sigma}F_{\rho\sigma}
    \right]
    + \ldots
    \label{eq:NLO_EOM1}
    \\
    \hspace{-.7cm}
    \os{G}{-4}_{h}^{\rho\sigma}
    &=h^{\mu\rho}h^{\nu\sigma}\left(
      \check R_{\mu\nu}
      - \frac{1}{2}h_{\mu\nu}h^{\kappa\lambda}\check R_{\kappa\lambda}
      - \left(\check{\nabla}_\mu + a_\mu\right)a_\nu 
      + h_{\mu\nu}h^{\kappa\lambda} \left(\check\nabla_{\kappa} + a_\kappa\right)a_\lambda
    \right)
    +\ldots
    \label{eq:NLO_EOM2}
  \end{align}
\end{subequations}
where the dots denote terms that vanish on shell upon using the $m_\mu$ and $\Phi_{\mu\nu}$ equations of motion, or equivalently upon imposing the TTNC condition.

Finally, the NNLO Lagrangian requires considerably more algebra.
Its form is simplest when we add a Lagrange multiplier to enforce the TTNC condition\footnote{The full result at NNLO without the TTNC condition can be found in \cite{Novosad:2021tlq}.}.
We refer to the result as the non-relativistic gravity (NRG) Lagrangian:
\begin{align}
  \mathcal{L}_\mathrm{NRG}
  &\equiv \left.\oss{\mathcal{L}}{-2}{NNLO}\right|_{\tau\wedge d\tau=0}+
    \frac{e}{16\pi G}\frac{1}{2}\zeta_{\rho\sigma}h^{\mu\rho}h^{\nu\sigma}(\partial_\mu\tau_\nu-\partial_\nu\tau_\mu)
    \nonumber\\
    &= \frac{e}{16\pi G}\Bigg[h^{\mu\rho}h^{\nu\sigma}K_{\mu\nu}K_{\rho\sigma}-\left(h^{\mu\nu}K_{\mu\nu}\right)^2
    -2m_\nu\left(h^{\mu\rho}h^{\nu\sigma}-h^{\mu\nu}h^{\rho\sigma}\right)\check{\nabla}_\mu K_{\rho\sigma}
    \nonumber\\
    &+\Phi h^{\mu\nu}\check R_{\mu\nu}
    +\frac{1}{4}h^{\mu\rho}h^{\nu\sigma}F_{\mu\nu}F_{\rho\sigma}+\frac{1}{2}\zeta_{\rho\sigma}h^{\mu\rho}h^{\nu\sigma}(\partial_\mu\tau_\nu-\partial_\nu\tau_\mu)
    \label{eq:action_NRG}
    \\
    &
    -\Phi_{\rho\sigma}h^{\mu\rho}h^{\nu\sigma}\left(\check R_{\mu\nu}-\check{\nabla}_\mu a_\nu - a_\mu a_\nu
      -\frac{1}{2}h_{\mu\nu}h^{\kappa\lambda}\check R_{\kappa\lambda}
  +h_{\mu\nu} e^{-1} \partial_{\kappa}\left(e h^{\kappa\lambda} a_\lambda\right)\right)\Bigg]\,,
  \nonumber
\end{align}
where $\Phi \equiv -v^\mu m_\mu$ is the Newtonian potential.
The resulting EOMs can be found in \cite{Hansen:2020pqs}. 

We remark that the NRG Lagrangian can also be obtained via another method.
This alternative route employs the type II TNC gauge symmetries and constructs the unique (up to a cosmological constant) two-derivative action respecting this symmetry, starting with the correct kinetic term required for Newton's law of gravitation and then completing the full action.
This was first done in \cite{Hansen:2018ofj} and elaborated on in \cite{Hansen:2020pqs}, and we refer to these papers for details, including the form of the EOMs.
As expected, the two Lagrangians are identical.
The main difference in their appearance arises from the fact that slightly different geometric variables are used in each, since the gauging construction naturally leads to manifestly Galilean boost invariant quantities. 
Depending on taste and type of application, one can work with either one of them. 

\subsection{Newton's Poisson equation}
\label{sec:NewtontypeII}
We will now show how the NRG action \eqref{eq:action_NRG} gives rise to the Poisson equation of Newtonian gravity when it is coupled to a massive non-relativistic particle.
In this way, we show that type II TNC geometry and the action~\eqref{eq:action_NRG} in particular provides an off-shell formulation of Newtonian gravity. 

We start by consider the large $c$ expansion of the Lorentzian particle Lagrangian 
\begin{equation}
  \label{eq:lor-pp-lag}
  \mathcal{L}=-mc\left(-g_{\mu\nu}\dot X^\mu\dot X^\nu\right)^{1/2}\,,
\end{equation}
where $X^\mu(\lambda)$ are embedding scalars and $\lambda$ is the geodesic parameter.
As before, we expand the metric according to \eqref{eq:tangent-metric-pnr-decomposition} and~\eqref{eq:pnr-down-expansion}.
In addition, we now also need to expand the embedding scalars according to the general expression~\eqref{eq:expansion_field_general}, which gives
\begin{equation}
  \label{eq:1c2_exp_embedding_scalar}
  X^\mu=x^\mu+\frac{1}{c^2}y^\mu+\order{c^{-4}}\,.
\end{equation}
This is necessary since the equations of motion for~$X^\mu$ would otherwise be overconstrained.
The resulting LO Lagrangian is then just
\begin{equation}
  \label{eq:pointparticle_LO_lagrangian}
  \oss{\mathcal{L}}{-2}{\text{LO}}=-m\tau_\mu\dot x^\mu\,,
\end{equation}
while the NLO Lagrangian becomes (up to a boundary term)
\begin{equation}
  \label{eq:NRparticle}
  \oss{\mathcal{L}}{0}{\text{NLO}}
  = m \left(
    \left(\partial_\nu\tau_\mu-\partial_\mu\tau_\nu\right)\dot x^\nu y^\mu
    +\frac{1}{2}\frac{\bar h_{\mu\nu}\dot x^\mu\dot x^\nu}{\tau_\rho\dot x^\rho}
  \right)\,.
\end{equation}
The latter is the Lagrangian of a particle on a type II TNC geometry.
Again, it can be checked that the LO EOM is correctly reproduced by the EOMs of $y^\mu$ in the subleading Lagrangian.

On a fixed torsionless NC background, the action~\eqref{eq:NRparticle} is the same as the standard point particle action~\eqref{eq:type-I-tnc-pp-action-from-null-red} on a torsionless type I TNC geometry \cite{Kuenzle:1972uk, Andringa:2010it, Bergshoeff:2015uaa}.
We also emphasize that LO term in the expansion of the particle action~\eqref{eq:lor-pp-lag} is of order $c^2$, so that it couples to the NNLO gravity action.
Likewise, the NLO particle action only couples to the N$^3$LO gravity action where it will source NNLO fields. This means that we can solve the geodesic equation at a given order in $1/c^2$ on a background whose fields were determined at the previous order. 
For example at order $c^2$ the geodesic equation, the equation of motion of $x^\mu$, tells us that $d\tau=0$. The coupling of $\oss{\mathcal{L}}{-2}{\text{LO}}$ to NRG leads to the equation of NC gravity as originally formulated by Trautman. We solve this equation for the NC variables and then at the next order $c^0$ we solve for the embedding scalars in the geodesic equation (Newton's law) obtained by varying $\oss{\mathcal{L}}{0}{\text{NLO}}$ with respect to $x^\mu$. This solution then sources the Einstein equations at order $c^0$, etc.

We thus consider the NRG Lagrangian~\eqref{eq:action_NRG} coupled to the LO point particle Lagrangian~\eqref{eq:pointparticle_LO_lagrangian}.
The $x^\mu$ equation enforces absolute time $d \tau =0$. In general, the equations of motion of the NRG part of the Lagrangian can 
be shown \cite{Hansen:2020pqs} to take the form 
\begin{equation}
  \label{eq:mattercoupledNCgravity}
  \bar R_{\mu\nu}
  = \frac{8\pi G}{d-1}\left(-(d-2)\tau_\rho \mathcal{T}_{\tau}^\rho
  + h_{\rho\sigma}\mathcal{T}_{h}^{\rho\sigma}\right)\tau_\mu\tau_\nu\,.
\end{equation}
where  $\mathcal{T}_\tau$ and $ \mathcal{T}_{h}$ are proportional to the responses of varying
the matter action with respect to $\tau_\mu$ and  $h_{\mu \nu}$ respectively, as will be reviewed
in more detail in Section~\ref{sec:matter}. 

In the case at hand, the sources that follow from the LO particle action are 
\begin{equation} 
  \mathcal{T}_\tau^\mu
  = - m\int d\lambda\,\frac{\delta(x-x(\lambda))}{e}\dot x^\mu \,,
\end{equation}
and $\mathcal{T}_h^{\mu\nu}=0$. 
The equations of motion of the NRG action~\eqref{eq:action_NRG} coupled to the point particle action~\eqref{eq:pointparticle_LO_lagrangian} then reproduce the covariant form of the Newtonian Poisson equation
\begin{equation}
  \label{eq:NewtontypeII}
  \bar R_{\mu\nu}= 8\pi G\frac{d-2}{d-1} \rho\, \tau_\mu \tau_\nu \,,
\end{equation}
with $\rho = m \int d\lambda\delta(x-x(\lambda))/e$ (in the gauge $\tau_\mu\dot x^\mu=1$). 
Besides these NC equations of motion, there are additional decoupled equations of motion for the field $\Phi_{\mu\nu}$ as well as the Lagrange multiplier. 
It is also important to stress that $\rho$ in \eqref{eq:NewtontypeII}  is not a Bargmann mass density but rather the leading contribution to the energy density. 
In all, we see that the NRG action~\eqref{eq:action_NRG} for type II TNC geometry provides a complete off-shell description of Newton--Cartan gravity with a point particle source.

\section{Other aspects of non-relativistic gravity}
\label{sec:aspectsNRG}
In this section, we discuss some further properties of the large speed of light expansion of general relativity.
This discussion is non-exhaustive and we will comment on other aspects and directions in Section~\ref{sec:discussion} below.

\subsection{Expansion of matter sources} 
\label{sec:matter}
We already discussed the coupling of a non-relativistic particle to non-relativistic gravity (NRG) in Section \ref{sec:NewtontypeII}.
Here, we present some highlights
of the large speed of light expansion
of more general matter couplings.
These expansions can be obtained by applying
the same methods as used for the Einstein–Hilbert (EH) Lagrangian
to a generic matter Lagrangian.
The matter actions that we obtain in this way act as sources for gravity in the $1/c^2$ expansion, and in particular for the NRG Lagrangian~\eqref{eq:action_NRG}.

The expansion of a generic matter Lagrangian takes the form
\begin{align}
  \hspace{-.2cm}
  \label{eq:expLagrangian_Matter}
  \mathcal{L}_{\text{mat}}(c^2,\phi,\partial_\mu\phi)
  &=c^{N}\tilde{\mathcal{L}}_{\text{mat}}(\sigma)
  \nonumber\\
  \hspace{-.2cm}
  &=c^{N}\;\oss{\mathcal{L}}{-N}{\text{mat},\,LO}
  +c^{N-2}\;\;\oss{\mathcal{L}}{2-N}{\text{mat},\,NLO}
  +c^{N-4}\;\;\oss{\mathcal{L}}{4-N}{\text{mat},\,NNLO}
  +\order{c^{N-6}}\,.
\end{align}
At each order $n\in\mathbb{Z}_{\geq0}$, we define matter currents as responses to variations of the geometric fields.
So for example, 
\begin{equation}
  \os{T}{2n-N}^{\alpha\beta}_{h}
  \equiv 2e^{-1}\frac{\delta\;\;{\os{\mathcal{L}}{2n-N}}_{\text{mat}\,,\mathrm{N}^n\mathrm{LO}}}
  {\delta h_{\alpha\beta}}\,,
  \label{eq:Matter_couplings_expansion}
  \qquad
  \os{T}{2n-N}^\alpha_{\tau}
  \equiv e^{-1}\frac{\delta\;\;{\os{\mathcal{L}}{2n-N}}_{\text{mat}\,,\mathrm{N}^n\mathrm{LO}}}
  {\delta \tau_{\alpha}}\,.
\end{equation}
In this way, one can get the EOMs of the sourced gravity coupled to matter at any order in
in the large $c$ expansion \cite{Hansen:2020pqs}.
In particular, if we define the EOMs from the variations of the $\mathrm{N}^n\mathrm{LO}$ gravity Lagrangian with respect to $h_{\alpha\beta}$ and $\tau_\alpha$ for $n\in \mathbb{Z}_{\geq0}$ as
\begin{equation}
  \label{eq:1c2_expansion_def_EOMs}
  \frac{1}{16\pi G}\;\os{G}{2n-6}^{\alpha\beta}_{h}
  \equiv -e^{-1}\frac{\delta\;\;{\os{\mathcal{L}}{2n-6}}_{\mathrm{N}^n\mathrm{LO}}}
  {\delta h_{\alpha\beta}}\,,
  \qquad
  \frac{1}{8\pi G}\;\os{G}{2n-6}^\alpha_{\tau}
  \equiv -e^{-1}\frac{\delta\;\;{\os{\mathcal{L}}{2n-6}}_{\mathrm{N}^n\mathrm{LO}}}
  {\delta \tau_{\alpha}}\,,
\end{equation}
we get the following sourced EOMs at any order $2m\geq-6$,
\begin{equation}
  \os{G}{2m}^{\alpha\beta}_{h}
  = 8\pi G\os{T}{2m}^{\alpha\beta}_{h} \,,
  \qquad
  \os{G}{2m}^{\alpha}_{\tau}
  = 8\pi G\os{T}{2m}^{\alpha}_{\tau} \,.
\end{equation}
The equations of motion and currents for the higher-order fields are defined analogously.
Relatedly, one can derive the Ward identities resulting from various gauge invariances, see \cite{Hansen:2020pqs} for details.
Amongst other things, these Ward identities provide the non-relativistic analogue of the conservation of the relativistic energy-momentum tensor.

A few remarks are in order. 
Recall that the LO EOMs $\os{G}{-6}^{\alpha\beta}_{h}$ and $\os{G}{-6}^{\alpha}_{\tau}$ are equivalent to $\tau\wedge d\tau=0$.
To avoid spacetimes which violate the twistless torsion condition $\tau\wedge d\tau=0$, we must therefore have matter actions such that the overall scaling $N$ in~\eqref{eq:expLagrangian_Matter} is at most equal to four, so that $\os{T}{-6}^{\alpha\beta}_{h}=\os{T}{-6}^{\alpha}_{\tau}=0$ and no violation of the twistless torsion condition is sourced.
It was shown in \cite{Hansen:2020pqs} that $N\leq 4$ for all examples considered, including the large $c$ expansion of a real or complex scalar field, the Maxwell field and fluids.
The non-relativistic point particle case discussed in subsection~\ref{sec:NewtontypeII}  corresponds to $N=2$. 

Thus, it is the matter sector that determines whether the geometry has torsion or not. 
This is seen for example in the expansion of perfect fluids, for which there are  different regimes depending on how we expand the energy and pressure as a function of $1/c^2$.
Ref. \cite{Hansen:2020pqs} furthermore discusses the resulting actions for
various field theory examples, a complex and real scalar field  as well as electrodynamics.
In particular, the expansion of a complex scalar field coupled to GR is shown to lead to the Lagrangian for the Schr\"odinger--Newton equation.
This off-shell description of the Schr\"odinger--Newton system includes fields whose equations of motion tell us that the clock one-form must be closed.
Furthermore, in the case of Maxwell theory, it is well-known that there are two limits (a magnetic and an electric limit, see~\cite{LeBellac,Duval:2014uoa,Festuccia:2016caf,Bergshoeff:2022qkx}) depending on how we expand the gauge connection, and one can obtain the Lagrangian descriptions for both using this procedure.

\subsection{Strong gravity expansion of the Schwarzschild metric}
\label{ssec:strong-grav-exp}
One way to generate solutions of non-relativistic gravity is by considering the $1/c^2$ expansion of solutions of GR. 
To illustrate this, we now discuss the $1/c^2$ expansion of the Schwarzschild solution.
Interestingly,
this can be done in two interesting ways, namely either through a weak field expansion related to the post-Newtonian expansion, or alternatively through a strong field expansion that leads to an exact torsionful solution of NRG.
While the precise physical interpretation of this latter expansion is still under construction, the fact that NRG includes solutions with torsion (so that time is no longer absolute) shows that it is richer than just Newtonian gravity. 

Consider the Schwarzschild metric including factors of $c$,
\begin{equation}
  \label{eq:schwarzschild_metric}
  d s^2
  = -c^2 \left(1-\frac{2G m}{c^2 r}\right)d t^2+\left(1-\frac{2Gm}{c^2 r}\right)^{-1} d r^2 + r^2 d\Omega_2^2\,.
\end{equation}
As first noticed in \cite{VandenBleeken:2017rij}, we can perform two different expansions, both of which are physically relevant, depending on how we treat the mass parameter $m$ as a function of~$c^2$.
The first option is to take $m$ to be constant in $c^2$ as we expand.
Using~\eqref{eq:tangent-metric-pnr-decomposition} and~\eqref{eq:pnr-down-expansion}, we see that the resulting the type II TNC fields are
\begin{subequations}
  \begin{align}
    \tau_\mu d x^\mu
    &= d t\,,
    \\
    m_\mu d x^\mu
    &= -\frac{Gm}{r}d t=\Phi d t\,,
    \\
    h_{\mu\nu}d x^\mu d x^\nu
    &= d r^2 + r^2d\Omega_2^2 \,,
    \\
    \Phi_{\mu\nu}d x^\mu d x^\nu
    &= \frac{2Gm}{r} d r^2 = -2\Phi d r^2\,.
  \end{align}
\end{subequations}
This is a flat torsionless Newton--Cartan spacetime with non-zero subleading fields $m_\mu$ and $\Phi_{\mu\nu}$.
It can be verified that this is a vacuum solution of the NRG equations of motion.
The solution has zero torsion and is expressed in terms of the Newtonian potential $\Phi=-v^\mu m_\mu=-Gm/r$. 
This geometry receives nontrivial corrections at subleading order.

In the second approach, we take the mass to be of order $c^2$, so that $M=m/c^2$ is constant in $c^2$, as was done in \cite{VandenBleeken:2017rij}.
This provides an approximation of GR that is distinct from the post-Newtonian expansion.
In this case, the expansion terminates at NLO, and the resulting geometry is described by the following type II TNC fields,
\begin{subequations}
  \begin{align}
    \tau_\mu d x^\mu
    &= \sqrt{1-\frac{2GM}{r}}d t\,,\label{eq:TTNCsol1}
    \\
    m_\mu d x^\mu
    &= 0\,,
    \\
    h_{\mu\nu}d x^\mu d x^\nu
    &= \left(1-\frac{2 GM}{r}\right)^{-1}d r^2 + r^2d\Omega_2^2 \,,
    \\
    \Phi_{\mu\nu}d x^\mu d x^\nu
    &= 0\,.\label{eq:TTNCsol4}
  \end{align}
\end{subequations}
This is a torsionful Newton--Cartan spacetime.
In fact, it is a vacuum solution of the equations of motion of the NLO Einstein--Hilbert Lagrangian \eqref{eq:NLOLag} which describes Galilean gravity, as it does not involve the subleading fields.
This strong gravity expansion of the Schwarzschild metric cannot be  captured by Newtonian gravity since it has nonzero torsion, but it can still be described as a torsionful Newton--Cartan geometry.
By studying the geodesics in this novel geometry, one can show
that the three classical tests of GR, namely perihelion precession, deflection of light and gravitational red-shift, are passed perfectly~\cite{Hansen:2019vqf}.
It is thus possible to have a non-relativistic causal structure
and still correctly include the effects of gravitational time dilation, albeit via a conceptually different way than in GR.

\subsubsection*{Other solutions}

We also mention here some other types of solutions that have been studied in the literature.
One can consider the Tolman-Oppenheimer-Volkoff fluid, where it turns out that the TOV equation can be derived entirely in the NR framework.
Another interesting case is that of cosmological solutions, where one can show that the FLRW spacetime is also an exact solution of NRG. 

One important remark is that the results of the $1/c^2$ expansion of a given Lorentzian spacetime can depend on the coordinate chart that is used to construct the expansion.
Given two different charts on a Lorentzian spacetime that are related by a diffeomorphism that is not analytic in $c$, the expansion of the spacetime in these charts will give rise to distinct non-relativistic spacetimes, which are not related by gauge transformations.
An example of this is the expansion of flat spacetime in the usual Minkowski coordinates versus Rindler coordinates, which lead to non-gauge-equivalent Newton--Cartan geometries.

As an additional example of the dependence of the expansion on the scaling we choose for the parameters in a solution, we consider two inequivalent $1/c^2$ expansions of $\mathrm{AdS}_{d+1}$.
In global coordinates and with explicit factors of $c$, the $\mathrm{AdS}_{d+1}$ metric is
\begin{equation}
  \label{eq:AdS_global}
  d s^{2}=-c^{2}\cosh^{2}\rho d t^{2}+l^2\left(d\rho^{2}+\sinh^{2}\rho d\Omega_{d-1}^{2}\right)\,,
\end{equation}
where $l$ is the AdS radius, $\rho>0$ is dimensionless and $t$ has dimensions of time.
The corresponding type II TNC geometry can be read off as  
\begin{subequations}
  \begin{align}
    \tau_{\mu}d x^{\mu}
    &= \cosh\rho d t\,,
    \\
    h_{\mu\nu}d x^{\mu}d x^{\nu}
    &= l^2\left(d\rho^{2}+\sinh^{2}\rho d\Omega_{d-1}^{2}\right)\,,
    \\
    m_\mu 
    &= 0\,,
    \\
    \Phi_{\mu\nu} 
    &= 0\,.
  \end{align}
\end{subequations}
and obviously the $1/c^{2}$ expansion terminates immediately. This is a torsionful NC spacetime.
Likewise, the $1/c^2$ expansion of AdS in Poincaré coordinates also leads to a torsionful type II TNC geometry.

On the other hand, if we do the coordinate transformation $r=l\sinh\rho$ in~\eqref{eq:AdS_global}, we obtain the metric 
\begin{equation}
  d s^2=-c^2\left(1+\frac{r^2}{l^2}\right)d t^2+\frac{d r^2}{1+\frac{r^2}{l^2}}+r^2d\Omega_{d-1}^2\,.
\end{equation}
Note that this metric describes the static patch of de Sitter if we replace $l^2$ by $-l^2$.
If we then define $l=\frac{c}{H}$ where $H$ is independent of $c$, we find
\begin{equation}
  ds^2=-c^2d t^2\mp H^2r^2d t^2+\frac{d r^2}{1\pm\frac{H^2r^2}{c^2}}+r^2d\Omega_{d-1}^2\,,
\end{equation}
where the upper sign is for AdS and the lower sign is for dS. 
Expanding  this to NLO, the resulting type II TNC geometry is
\begin{equation}\label{eq:NHspaces}
    \tau=dt\,
    \qquad
    h_{\mu\nu}d x^\mu dx^\nu
    = d\vec x\cdot d\vec x\,,
    \qquad
    m_\mu dx^\mu
    = \pm \frac{1}{2}H^2 \vec x^2 dt\,,
\end{equation}
where $\Phi_{\mu\nu}$ is omitted and we transformed to Cartesian coordinates. This TNC geometry is known as the Newton--Hooke spacetime.
In \cite{Grosvenor:2017dfs} it was shown that such a spacetime can be written in the form of a non-relativistic FLRW geometry with flat spatial slices by a sequence of NC gauge transformations. 
Furthermore, they were related to a null reduction of a pp-wave geometry in~\cite{Gibbons:2003rv}.

\subsection{Odd powers in \texorpdfstring{$1/c$}{1/c}}
\label{ssec:odd}
So far, we have focused on even powers of $1/c$ in our large $c$ expansion.
As we have seen, this is a consistent subsector in the purely gravitational sector.
In earlier work~\cite{Dautcourt:1996pm},
using a weak field assumption along with physical constrains on the energy-momentum tensor,
it was shown that odd terms only appear at subleading orders, beyond the 1PN order in post-Newtonian expansions. 
A more complete analysis of odd powers, including ones that can appear at pre-Newtonian orders and going beyond the weak field assumption, was performed in~\cite{Ergen:2020yop}.
The motivation for this is that an energy-momentum tensor that sources torsion can possibly also source the leading-order odd term in the metric at order $c^1$ when there is dynamics.
Another motivation is that a solution such as the Kerr metric admits several $1/c$ expansions depending on how one scales the mass and angular momentum, and some of these lead to odd powers in $1/c$~\cite{Ergen:2020yop}. Finally, odd powers in $1/c$ can capture retardation effects.
 
The starting point of \cite{Ergen:2020yop} follows from writing the line element as follows,
\begin{equation}
  \begin{split}
    ds^2
    &= -e^{-\Psi}  (cdt + C_i dx^i)^2
    +e^{\Psi}k_{ij}dx^i dx^j
    \\
    &{}\qquad
    + \mathcal{O} (c) dt^2
    + \mathcal{O}_i (c^0)dtdx^i
    + \mathcal{O}_{ij}(c^{-1})dx^idx^j\,.
  \end{split}
\end{equation}
The authors then obtain the LO equations of motion of the LO physical fields, which consist of 
a scalar potential $\Psi$, a vector potential $C_i$ and a spatial metric $k_{ij}$.  
When these fields are time-independent, they can be shown to be solutions
to the Einstein equations for stationary metrics.
The interesting observation is that when we do allow for time dependence, the LO fields still satisfy the same equations as when no time dependence is involved.
This means that the time dependence sits in the integration `constants' when solving the Einstein equations for stationary metrics. These time-dependent integration constants source the next subleading equations.
We can thus view the $1/c$ expansion as an expansion around a stationary GR solution.
This has been illustrated for the Kerr metric in~\cite{Ergen:2020yop}.
In a similar fashion, the $1/c^2$ expansion can be viewed as an expansion around a static sector of GR \cite{VandenBleeken:2017rij}.

\section{Discussion}\label{sec:discussion}
In this review we have put the  main emphasis on introducing the reader to the various notions of Newton-Cartan geometry
and how they enter the non-relativistic expansions of general relativity.
We only mentioned a few applications,
but there are numerous further developments and extensions related to the topic of this review, which we briefly mention here.

\paragraph{Connection to the Post-Newtonian expansion.}
The post-Newtonian expansion is performed in harmonic gauge. The Einstein equations can be formally integrated using a retarded Green's function (to obey a no-incoming-radiation boundary condition in the past).
One then solves this integral equation by performing both a $1/c$ expansion (in a large but finite region containing the source) and a $G$ expansion (outside the source) and matching the two in their overlap region.
However, there currently does not exist a calculational scheme that allows one to do this in an arbitrary gauge. 

The covariant $1/c$ expansion seems the ideal starting point to try to generalise the existing approaches such as the ones due to Blanchet--Damour and Will--Wiseman.
However, this would be very reminiscent of what is sometimes called the `classic' approach.
The latter has been abandoned because the $1/c$ expansion beyond the 1PN order has a finite regime of validity and so one cannot impose asymptotic boundary conditions on the $1/c$ solution.
Nevertheless, we want to advocate a hybrid approach that combines the Blanchet--Damour or the Will--Wiseman approach with the classic approach.
We refer to \cite{HartongMusaeus} for more details.

\paragraph{Carroll expansion of gravity.}
The study of the small speed of light limit and expansion of GR goes back to~\cite{Henneaux:1979vn} and later \cite{Dautcourt:1997hb}. This expansion is also called an ultra-local or Carroll expansion, since the Poincar\'e group contracts to the Carroll group  \cite{Levy1965,Bacry:1968zf,SenGupta} in the $c \rightarrow 0$ limit. The systematic study of the small speed of light expansion of GR, paralleling the approach of \cite{Hansen:2020pqs}, was recently obtained in \cite{Hansen:2021fxi}, to which we refer for further references on Carroll geometry and Carrollian gravity theories. It is interesting to note that while the LO action in the large $c$ expansion is just the TTNC condition, in the Carroll expansion the LO action already involves non-trivial (though ultra-local) dynamics.

\paragraph{Other formulations.} 
Different approaches to frame and/or first-order formulations of non-relativistic (and ultra-local) limits and expansions of gravity have been considered in 
\cite{Bergshoeff:2017btm,Cariglia:2018hyr,Guerrieri:2020vhp,Hansen:2021fxi}. 
Furthermore, in Ref.~\cite{Hansen:2020wqw} the Palatini action for GR was reformulated in terms of moving frames that exhibit local Galilean covariance in a large speed of light expansion.  

There are also interesting connections between the 1+3 formulation used in GR and the non-relativistic expansion. A 1+3 formulation of Newton's equations was discussed in \cite{DePietri:1994je} and in \cite{Vigneron:2020rwy}
and applied to cosmology in \cite{Vigneron:2020ewy}. 
The relation to the 1+3 formulation was  performed in a more systematic fashion in  \cite{Elbistan:2022plu}, extending the computation of the effective Lagrangian to
a higher order and making some new all-order observations.

\paragraph{NR gravity models in two and three spacetime dimensions.} 
Just as for GR, special types of models exist when considering non-relativistic gravity in two and three spacetime dimensions.
Three-dimensional CS theories of non-relativistic gravity based on 
extended Galilei algebras were first obtained in~\cite{Papageorgiou:2009zc,Bergshoeff:2016lwr,Hartong:2016yrf} and further studied and generalized in e.g. \cite{Gomis:2019nih,Matulich:2019cdo,Concha:2020ebl,Concha:2020eam,Concha:2022you}.
 Likewise non-relativistic (and Carrollian) versions of JT gravity were
 first given in \cite{Grumiller:2020elf,Gomis:2020wxp} and generalized
 in  \cite{Ravera:2022buz}.

\paragraph{Non-relativistic string theory.} 
An entire subject on its own, while also closely related to that of the present review, is the topic of non-relativistic string theory.
For this we refer the reader to the recent review article \cite{Oling:2022fft}, which together with the present review article is part of a larger set of reviews related to modern developments in non-relativistic physics. 
Non-relativistic string theory on flat spacetime was first formulated in \cite{Gomis:2000bd}. A natural question, which has only been recently addressed motivated in large part by the developments reviewed here, is the question what type of curved target spacetime geometry non-relativistic strings move in. This study was initiated in \cite{Andringa:2012uz,Harmark:2017rpg,Bergshoeff:2018yvt} and we refer to the review \cite{Oling:2022fft} for a more complete set of references. See also the recent Ref.~\cite{Bidussi:2021ujm} for a derivation of type I torsional string Newton-Cartan (TSNC) geometry along similar lines as those 
presented in Section~\ref{sec:type-I-tnc} of this review, using both the perspective of the limit and null reduction
of relativistic string actions, as well as the gauging of the so-called fundamental string Galilei algebra.
Moreover, the stringy analogue of type II TNC, called type II TSNC, was recently discussed in \cite{Hartong:2021ekg,Hartong:2022dsx}. 
We also note that when quantizing non-relativistic string theory, one finds low energy effective actions
\cite{Gomis:2019zyu,Gallegos:2019icg,Bergshoeff:2019pij,Bergshoeff:2021bmc} that are similar to the type I TNC actions described in this review, though they involve different geometric fields.

\paragraph{Non-relativistic holography.}  
Non-relativistic, or more generally non-Lorentzian, geometry and gravity also play a role in non-AdS holography. In its original form, the AdS/CFT correspondence relates a relativistic bulk gravity to a corresponding dual (conformal) relativistic field theory living on the boundary. Beyond this, roughly three other classes of dualities have been found involving some type of non-Lorentzian geometry.  The first is the appearance of non-Lorentzian geometry on the
boundary.
This was uncovered in the context of Lifshitz holography, 
which led to the discovery of a torsionful generalisation
of NC geometry \cite{Christensen:2013lma,Christensen:2013rfa,Hartong:2014pma}. 
The reason that Newton--Cartan geometry appears is that light cones open up as one approaches the boundary.
This observation spurred on many of the subsequent developments in non-relativistic geometry.

Additionally, it has been suggested that non-relativistic field theories have perhaps a more natural holographic realization with non-relativistic gravity theories in the bulk 
\cite{Kachru:2008yh,Griffin:2012qx,Hofman:2014loa,Hartong:2016yrf}. 
This also seems to be the case for the holographic bulk duals of Spin Matrix Theory \cite{Harmark:2014mpa}, which are quantum-mechanical theories obtained from near-BPS limits of AdS${}_5$/CFT${}_4$.
On the string theory side, these are described by novel non-relativistic worldsheet models \cite{Harmark:2017rpg,Harmark:2018cdl,Harmark:2019upf,Harmark:2020vll,Kluson:2021sym}, which in the low-energy limit are expected to be described by dynamical non-Lorentzian gravity theories. 
Finally, we mention that there also exists an example \cite{Hartong:2017bwq} of a non-relativistic bulk gravity theory with a scale invariant relativistic field theory on the boundary.

\paragraph{Supersymmetry.} 
A natural generalization to consider is supersymmetric extensions of non-relativistic gravity.
This is especially relevant in view of the connection of NC gravity with string theory and holography as well as using supersymmetric localization techniques for non-relativistic field theories. 
We refer to the reader to the recent review article~\cite{Bergshoeff:2022iyb}, which is also connected to the present review. 
In that review, an overview is given of the different non-Lorentzian supergravity theories in diverse dimensions that have been constructed in recent years. 

\paragraph{Generalizations.}

We also mention here a number of further generalizations involving some type of non-relativistic gravity. Ref.~\cite{Hartong:2015zia} showed that TNC geometry is a natural geometrical framework underlying Ho\v{r}ava--Lifshitz gravity with manifest diffeomorphism invariance.
This connection was further studied in \cite{Afshar:2015aku,Hartong:2016yrf}. 
A teleparallel version of NC gravity was considered 
 in \cite{Read:2018ogw,Schwartz:2022bgz}, while a non-relativistic MacDowell-Mansouri
type approach was considered in \cite{Concha:2022jdc}. A generalization 
of NRG for arbitrary co-dimension foliation was presented in \cite{Novosad:2021tlq}.
Two further generalizations include a non-relativistic version of spin-3 CS gravity \cite{Bergshoeff:2016soe,Concha:2022muu} as well as  multi-metric gravity \cite{Ekiz:2022wbi}. 

\paragraph{Field theory applications.}
Another point worth mentioning is that TNC geometry plays an important role as the natural  background geometry \cite{Son:2013rqa,Jensen:2014aia,Hartong:2014oma,Geracie:2014nka} in non-relativistic field theories, which are ubiquitous in condensed matter and biological systems.  Further applications of this involving  energy-momentum tensors, Ward identities,  hydrodynamics and anomalies in the context of non-relativistic field theories can be found e.g. in  \cite{Jensen:2014ama,Geracie:2015xfa,Hartong:2016nyx,deBoer:2017ing,deBoer:2017abi,Armas:2019gnb}). 
Last but not least, the use of type II TNC is expected to be important to further examine the physics of non-relativistic quantum matter in a non-trivial  non-relativistic  spacetime geometry.  Developing this framework  could allow to address new signals at low energies for quantum matter in gravitational backgrounds and study composite systems with potentially measurable decoherence effects.

\vspace{\baselineskip}

In all, we conclude with the observation that the field of non-relativistic and more generally non-Lorentzian gravity and its relation to field theory, gravity and string theory is still growing in scope.
Each of the exciting research lines mentioned above is expected to develop further in the coming years. 

\subsection*{Acknowledgements}

We thank Emil Have, J{\o}rgen S.\ Musaeus and Benjamin T.\ S{\o}gaard for useful discussions and Dennis Hansen for collaboration on many of the results reviewed here. 
The work of JH is supported by the Royal Society University Research Fellowship ``Non-Lorentzian Geometry in Holography'' (grant number UF160197). 
The work of NO is supported in part by the project ``Towards a deeper understanding of  black holes with non-relativistic holography'' of the Independent Research Fund Denmark (grant number DFF-6108-00340).
The work of NO and GO is furthermore supported by the Villum Foundation Experiment project 00023086 and by the VR project grant 2021-04013. 

\appendix

\section{Conventions}\label{sec:conventions}

For a connection with torsion whose connection coefficients are $\Gamma^\rho_{\mu\nu}$ the Riemann tensor ${R}_{\mu\nu\sigma}{}^{\rho}$  and torsion tensor $T^\rho{}_{\mu\nu}$ are given by
\begin{eqnarray}
\left[\nabla_\mu,\nabla_\nu\right]X_\sigma & = & R_{\mu\nu\sigma}{}^\rho X_\rho-T^\rho{}_{\mu\nu}\nabla_\rho X_\sigma\,,\\
\left[\nabla_\mu,\nabla_\nu\right]X^\rho & = & -R_{\mu\nu\sigma}{}^\rho X^\sigma-T^\sigma{}_{\mu\nu}\nabla_\sigma X^\rho\,,
\end{eqnarray}
which implies
\begin{eqnarray}
{R}_{\mu\nu\sigma}{}^{\rho}&\equiv&-\partial_{\mu}{\Gamma}_{\nu\sigma}^{\rho}+\partial_{\nu}{\Gamma}_{\mu\sigma}^{\rho}-{\Gamma}_{\mu\lambda}^{\rho}{\Gamma}_{\nu\sigma}^{\lambda}+{\Gamma}_{\nu\lambda}^{\rho}{\Gamma}_{\mu\sigma}^{\lambda}\,,\\
T^\rho{}_{\mu\nu} &\equiv& 2\Gamma^\rho_{[\mu\nu]}\,.
\end{eqnarray}
These obey the following Bianchi identities
\begin{eqnarray}
    R_{[\mu\nu\sigma]}{}^\rho & = & T^\lambda{}_{[\mu\nu}T^\rho{}_{\sigma]\lambda}-\nabla_{[\mu}T^\rho{}_{\nu\sigma]}\,,\\
    \nabla_{[\lambda}R_{\mu\nu]\sigma}{}^\kappa & = & T^\rho{}_{[\lambda\mu}R_{\nu]\rho\sigma}{}^\kappa\,.
\end{eqnarray}
We define the Ricci tensor as
\begin{equation}\label{eq:Ricci_tensor_LC}
R_{\mu\nu} \equiv {R}_{\mu\rho\nu}{}^{\rho}\,.
\end{equation}
We will assume that the connection is such that 
\begin{equation}
    \Gamma^{\rho}_{\mu\rho}=\partial_\mu\log M\,,
\end{equation}
where $M$ is the integration measure, so that
\begin{equation}
R_{\mu\nu\rho}{}^\rho=0\,.
\end{equation}

\section{Details for Trautman condition computation}\label{app:Trautman}

In Section~\ref{subsec:Trautman} it is left to show that $h^{\mu[\gamma}\check R_{\mu(\nu\sigma)}{}^{\rho]}=0$. This expression can be shown to be Galilean boost invariant. We prove $h^{\mu[\gamma}\check R_{\mu(\nu\sigma)}{}^{\rho]}=0$ by showing that all the projections with $v^\nu v^\sigma$, $v^\nu h^{\sigma\lambda}$ and $h^{\nu\kappa}h^{\sigma\lambda}$ give zero. We will use that 
    \begin{equation}\label{eq:K}
        \check\nabla_\rho v^\mu=-h^{\mu\nu}K_{\rho\nu}\,,
    \end{equation}
    where $K_{\mu\nu}=-\frac{1}{2}\mathcal{L}_v h_{\mu\nu}$ with $\mathcal{L}_v$ the Lie derivative along $v^\mu$. Using the definition of the Riemann tensor as well as \eqref{eq:K} it follows that
    \begin{equation}
        v^\nu v^\sigma\check R_{\mu\nu\sigma}{}^{\rho}=-h^{\rho\lambda}v^\kappa\check\nabla_\kappa K_{\mu\lambda}\,.
    \end{equation}
    Hence we find $h^{\mu[\gamma}\check R_{\mu(\nu\sigma)}{}^{\rho]}v^\nu v^\sigma=0$ which holds since $h^{\mu\nu}$ is covariantly constant and we also used the fact that $K_{\mu\nu}$ is symmetric.
    We next turn to the projection of $h^{\mu[\gamma}\check R_{\mu(\nu\sigma)}{}^{\rho]}=0$ with $h^{\nu\kappa}h^{\sigma\lambda}$. To this end define
    \begin{equation}
    X^{\alpha\beta\mu\nu}=h^{\alpha\rho}h^{\beta\sigma}\check R_{\rho\sigma\kappa}{}^\nu h^{\mu\kappa}\,.
    \end{equation}
    The tensor $X^{\alpha\beta\mu\nu}$ is antisymmetric in its first two and last two indices\footnote{Antisymmetry in the last two indices follows from $[\check\nabla_\rho\,,\check\nabla_\sigma]h^{\mu\nu}=0$ which leads to $\check R_{\rho\sigma\kappa}{}^{(\nu}h^{\mu)\kappa}=0$.}. It also obeys $X^{[\alpha\beta\mu]\nu}=0$. Adding and subtracting $X^{[\alpha\beta\mu]\nu}=0$, $X^{[\alpha\beta\nu]\mu}=0$, $X^{[\beta\mu\nu]\alpha}=0$ and $X^{[\mu\nu\alpha]\beta}=0$ appropriately we obtain $X^{\alpha\beta\mu\nu}=X^{\mu\nu\alpha\beta}$. This can be used to show that $X^{\mu(\nu\alpha)\beta}-X^{\beta(\nu\alpha)\mu}=0$ which is equivalent to $h^{\rho[\alpha}\check R_{\rho(\sigma\kappa)}{}^{\nu]}h^{\beta\sigma}h^{\mu\kappa}=0$.

Finally we need to show that $h^{\rho[\alpha}\check R_{\rho(\sigma\kappa)}{}^{\nu]}v^{\sigma}h^{\mu\kappa}=0$. To this end define
\begin{equation}
    Y^{\alpha\mu\nu}=h^{\rho\alpha}\check R_{\rho\sigma\kappa}{}^\nu v^\sigma h^{\kappa\mu}\,.
\end{equation}
In terms of this new object we have
\begin{equation}
    4h^{\rho[\alpha}\check R_{\rho(\sigma\kappa)}{}^{\nu]}v^{\sigma}h^{\mu\kappa}=Y^{\alpha\mu\nu}-Y^{\nu\mu\alpha}+2h^{\rho[\alpha}\check R_{\rho\kappa\sigma}{}^{\nu]}v^{\sigma}h^{\mu\kappa}\,.
\end{equation}
Using $\check R_{[\rho\sigma\kappa]}{}^\nu=0$ we can show that
\begin{equation}\label{eq:intermed}
    Y^{\alpha\mu\nu}-Y^{\mu\alpha\nu}+h^{\rho\alpha}\check R_{\kappa\rho\sigma}{}^{\nu}v^{\sigma}h^{\mu\kappa}=0\,.
\end{equation}
Cyclically permuting the indices on this last equation leads to two more equations. Adding and subtracting these off \eqref{eq:intermed} leads to 
\begin{equation}
    2Y^{\alpha\mu\nu}=-h^{\rho\alpha}\check R_{\kappa\rho\sigma}{}^{\nu}v^{\sigma}h^{\mu\kappa}-h^{\rho\nu}\check R_{\kappa\rho\sigma}{}^{\mu}v^{\sigma}h^{\kappa\alpha}+h^{\rho\mu}\check R_{\kappa\rho\sigma}{}^{\alpha}v^{\sigma}h^{\kappa\nu}\,.
\end{equation}
This tells us that
\begin{equation}
    Y^{\alpha\mu\nu}-Y^{\nu\mu\alpha}=-h^{\rho\nu}\check R_{\kappa\rho\sigma}{}^{\mu}v^{\sigma}h^{\kappa\alpha}\,.
\end{equation}
This in turn can be used to obtain
\begin{equation}
    4h^{\rho[\alpha}\check R_{\rho(\sigma\kappa)}{}^{\nu]}v^{\sigma}h^{\mu\kappa}=-h^{\rho\nu}\check R_{\kappa\rho\sigma}{}^{\mu}v^{\sigma}h^{\kappa\alpha}+h^{\rho\alpha}\check R_{\rho\kappa\sigma}{}^{\nu}v^{\sigma}h^{\kappa\mu}-h^{\rho\nu}\check R_{\rho\kappa\sigma}{}^{\alpha}v^{\sigma}h^{\kappa\mu}\,.
\end{equation}
Finally, using 
\begin{equation}
    \check R_{\kappa\rho\sigma}{}^{\mu}v^{\sigma}=h^{\mu\gamma}\left(\check\nabla_\kappa K_{\rho\gamma}-\check\nabla_\rho K_{\kappa\gamma}\right)\,,
\end{equation}
which follows from the definition of the Riemann tensor and \eqref{eq:K}, we obtain the result that
\begin{equation}
    h^{\rho[\alpha}\check R_{\rho(\sigma\kappa)}{}^{\nu]}v^{\sigma}h^{\mu\kappa}=0\,.
\end{equation}

\addcontentsline{toc}{section}{References}
\bibliographystyle{JHEP}
\bibliography{refs_NRG}

\end{document}

%% file: defs.tex
\newcommand{\pd}{\partial}

\newcommand{\LL}{\mathcal{L}}

\newcommand{\OO}{\mathcal{O}}
\newcommand{\qiq}{\quad\implies\quad}

